\documentclass[pra,twocolumn,showpacs,amsmath,amssymb,superscriptaddress,unsortedaddress,floatfix,longbibliography]{revtex4-1}
\usepackage{graphicx}
\usepackage{latexsym}
\usepackage{amsmath}
\usepackage{graphics}
\usepackage{amssymb}
\usepackage{layout}
\usepackage{verbatim}
\usepackage{amsfonts,epsfig,dsfont}
\usepackage{natbib}
\usepackage{hyperref}
\usepackage{color}
\usepackage[latin1]{inputenc}       
\usepackage[T1]{fontenc}

\newcommand{\ket}[2][]{{|#2\rangle_{#1}}}

\def\duzomniejsze{<\kern-.7mm<}
\def\duzowieksze{>\kern-.7mm>}

\def\textbf#1{{\bf #1}}
\def\beq{\begin{equation}}
\def\eeq{\end{equation}}
\def\be{\begin{equation}}
\def\ee{\end{equation}}
\def\ben{\begin{eqnarray}}
\def\een{\end{eqnarray}}
\def\beqa{\begin{eqnarray}}
\def\eeqa{\end{eqnarray}}
\def\eea{\end{array}}
\def\bea{\begin{array}}
	\newcommand{\unit}{\mathbb{I}}

	\usepackage{color}
	
	\usepackage{ulem} % Crossing out with \sout{...}
	
\begin{document}
	
	\title{Equivalence of qubit-environment entanglement and discord generation via pure dephasing interactions and the consequences thereof} 
	
	\author{Katarzyna Roszak}
	\affiliation{Department of Theoretical Physics, Faculty of Fundamental Problems of Technology, Wroc{\l}aw University of Science and Technology,
		50-370 Wroc{\l}aw, Poland}
	\affiliation{Institute of Physics, Academy of Sciences of the Czech Republic, 18221 Prague, Czech Republic}
	
	\author{{\L}ukasz Cywi{\'n}ski}
	\affiliation{Institute of Physics, Polish Academy of Sciences, Aleja Lotnikow 32/46, 02-668 Warsaw, Poland}
	
	\date{\today}

	\begin{abstract}
		We find that when a qubit initialized in  a pure state experiences pure dephasing due to interaction with an environment, separable qubit-environment states generated during the evolution also have zero quantum discord with respect to the environment.
	    What follows is that 
		the set of separable states which can be reached during the evolution
		has zero volume and hence, such effects as sudden death of qubit-environment entanglement are
		very unlikely. In case of the discord with respect to the qubit, a vast 
		majority of separable states qubit-environment is discordant, but in specific situations
		zero-discord states are possible. This is conceptually important since there is a connection
		between the discordance with respect to a given subsystem and the 
		possibility of describing the evolution of this subsystem
		using completely positive maps.
		Finally, we use the formalism to find an exemplary evolution of an entangled state of two qubits that is completely positive, occurs solely due to interaction of only one of the qubits with its environment (so one could guess that it corresponds to a local operation, since it is local in a physical sense), but which nevertheless causes the enhancement of entanglement between the qubits. While this simply means that the considered evolution is completely positive, but does not belong to LOCC, it shows how much caution has to be exercised when identifying  evolution channels that belong to that class.
	\end{abstract}

	\maketitle

	\section{Introduction}
	
	There is little or no ambiguity in the study of quantum correlations for pure
	states, as long as the potentially correlated parties are well defined
	and completely distinguishable. Such correlations can be fully described by
	entanglement and none of the pure separable states (states with no entanglement) exhibit any type of
	behaviors which can be associated with quantum correlations.
	The two main characteristics of pure entangled states are that
	(a) it is not possible to prepare an entangled state via local operations
	and classical communication (criterion of preparation) and (b) it is not possible to fully determine the 
	state of either of the entangled subsystems by local measurements on this subsystem and classical communication alone
	without disturbing it (criterion of measurement). It is this second characteristic 
	which results in the property of entangled states that appropriately chosen
	measurements on one subsystem determine the state of the other subsystem,
	which underlie many applications of entangled states such as quantum algorithms
	\cite{deutsch92,Ekert_RMP96}
	or quantum teleportation \cite{Bennett_PRL93,Popescu_PRL94}. 
	
	In case of mixed states, the situation becomes more complicated. Mixed state 
	entanglement \cite{werner89,Plenio_QIC07,horodecki09} is defined using the criterion of preparation of the previous paragraph, meaning
	that a state of two subsystems is entangled, if and only if it cannot be written as a statistical mixture
	of product states of the two subsystems -- equivalently, if it cannot be prepared by local operations and classical communication (LOCC).
	All entangled states satisfy 
	the criterion of measurement as well, but there exist separable (not entangled) states
	which satisfy the criterion of measurement for one or both subsystems (while obviously not satisfying
	the criterion of preparation). A measure of quantum correlations which is based
	on the criterion of measurement is called the quantum discord \cite{ollivier01,henderson01,modi14}. 
	The set of discordant states is larger
	than the set of entangled states and in fact, it includes the set of entangled 
	states, so although there do not exist entangled states with zero discord, there
	do exist separable discordant states \cite{modi14}. 
	It is important to note that there is an inherent asymmetry in the definition
	of the quantum discord with respect to the potentially correlated subsystems,
	since the criterion of measurement can be fulfilled for one of the subsystems
	while it is not fulfilled for the other. 
	
	Entanglement and the quantum discord differ significantly when it comes
	to the qualitative and quantitative features of their evolution. This is partly
	because the set of zero-discord states has zero volume \cite{ferraro10},
	while the volume of separable states is finite. The characteristic property
	for entanglement evolutions, the possibility for it to undergo sudden death
	\cite{rajagopal01,zyczkowski01,yu06}
	(the complete disappearance of entanglement at a certain finite time, 
	while the continuous decoherence of the entangled subsystems is not complete),
	which is sometimes followed by sudden birth 
	(the reemergence of entanglement after a state is separable
	for a finite amount of time), is a direct result of the geometry of separable
	states. Since they have finite volume, there exist separable states which are 
	completely surrounded by other separable states and may not be approached 
	by means of a continuous evolution otherwise than from another separable state.
	Any evolution which approaches such a state has to display sudden death of 
	entanglement.
	Since the volume of zero-discord states is not finite, any zero-discord state
	can be reached directly from a discordant state and
	sudden-death-type
	behavior in the discord evolution is much less likely. 
	
	On the other hand,
	entanglement is symmetric with respect to both entangled subsystems, while the 
	discord does not have to be symmetric with respect to the systems under study
	\cite{modi14}
	(it is fairly common that the measurement of one of the correlated subsystems
	yields information about its state with less damage to the state itself
	than the other - some of the geometric measures are artificially symmetrized
	\cite{dakic10}, or that a state is discordant only with respect to one of 
	the subsystems). 
	Another characteristic property of discord evolutions
	is the occurrence of points of indifferentiability (for which the time-dependence
	of the discord function is continuous, but not smooth). This quality of the
	discord should not be dismissed as an artifact of the mathematical properties of the geometric measures used
	to quantify the discord, since it has been  
	observed in quantum discord curves calculated using the original discord definition
	\cite{roszak15b,mazzola10}
	and in case of states which do have parity symmetry \cite{roszak13,mazurek14}.
	
	We focus here on quantum correlations, especially quantum discord, that appear between the system (a qubit or a pair of qubits) and its environment in the course of decoherence of the system. Specifically, we consider here the system initialized in a pure state (obviously completely uncorrelated with the environment), that interacts with the environment via pure-dephasing type coupling that singles out a basis of pointer states, and we consider states of the system that are superpositions of them. Generation of qubit-environment entanglement in this case was a subject of previous works \cite{Eisert_PRL02,Hilt_PRA09,Maziero_PRA10,Pernice_PRA11,roszak15a,Pozzobom_AP17}, and here, buliding on  the results of \cite{roszak15a}, we investigate the generation of system-environment discord (both with respect to the environment and to the qubit) during such evolution.  
	In the following, we show that strong quantum correlations described by
	entanglement and weaker quantum correlations described by the quantum discord
	with respect to the environment
	are operationally the same. This means that the class of separable
	qubit-environment (Q-E) states 
	that can be reached during such joint evolution, is the same as the class of 
	reachable zero-discord states {\it from the point of view of E}. Hence, all separable Q-E states obtained during the evolution are automatically one-sidedly zero-discord states, so they posses the specifically non-discordant
	quality of being a zero-volume set of states. This suppresses the possibility
	for such Q-E evolutions to display characteristic behaviors for
	entanglement, such as its sudden death. 
	
	On the other hand, when we look at the property of discordance with respect to the system, it turns out to have qualitatively different properties than the discord
	discussed in the previous paragraph. In fact most of the separable qubit-environment
	evolutions are discordant in this sense, and only in very specific situations
	are zero-discord points possible during the evolution. This means that in terms
	of weak quantum correlations, two types of evolutions are possible.
	
	After achieving such an understanding of Q-E discord generation, in the last part of the paper we use these results to shed light on issues related to role of system-environment quantum discord in dynamics of open systems.  We generalize our results on Q-E evolution to the case of a class of entangled two-qubit states subjected to pure dephasing due to $E$. Then we construct an example of a system in which only one qubit interacts with $E$, so that the resulting decoherence is local, and use the evolution due to the interaction with $E$ to find an example of a state $\hat{\rho}_{\mathrm{SE}}^{\mathrm{min}}$ with zero discord between the qubits and $E$ (zero with respect to the two-qubit subsystem), for which entanglement between the qubits is minimal, but subsequent evolution leads to an {\it increase} of interqubit entanglement. It is known that  the lack of system-environment discord with respect to the system implies that the system's evolution starting from such a state may be described using completely positive (CP) maps \cite{Rodriguez_JPA08}, so the qubits' evolution starting from $\hat{\rho}_{\mathrm{SE}}^{\mathrm{min}}$  state of the whole system is CP, but, despite the fact that only one of the qubits is interacting with its local environment, it is does not belong to the LOCC class. 
	
	The paper is organized as follows. In Sec.~\ref{sec2} we introduce the notion
	of the quantum discord further, including the original definition of the discord,
	and focusing on the differences and similarities between separable states and zero-discord states. In Sec.~\ref{sec3} we state the criteria for zero-discord
	states following Ref.~\cite{huang11}, which we will later use to obtain the main
	results of this paper. The class of systems under study is described in 
	Sec.~\ref{sec4}, and the separability criterion specific for this system
	is the topic of Sec.~\ref{sec5}. The equivalence of the class of separable
	states and zero-discord states with respect to the environment
	for the class of systems under study is 
	shown in Sec.~\ref{sec6}.
	The properties of the quantum discord with respect to the qubit are discussed
	in Sec.~\ref{sec6a}, while the an extension of the results to entangled 
	two-qubit states is the topic of Sec.~\ref{sec6b}.
	In Sec.~\ref{sec7} we discuss the implications of our results for the understanding of open quantum systems dynamics.
	
	\section{Quantum discord \label{sec2}}
	
	The quantification of the quantum discord is in general complicated 
	\cite{Huang_NJP14}, even 
	in comparison with the stronger measure of quantum correlations, entanglement.
	In case of entangled mixed states, some means of quantification of the amount
	of mixed state entanglement have been available for two decades. This includes entanglement
	witnesses, a multitude of two-qubit measures \cite{wootters98,vedral98,horodecki01,Plenio_QIC07}, which allow for the calculation
	of two-qubit or qubit-qutrit entanglement directly from the density matrix.
	
	Contrarily, the first geometric measures of the quantum discord
	(measures based on the calculation of the smallest distance between a given
	mixed quantum state and the set of zero-discord states; the distance
	measures used vary) 
	\cite{dakic10,luo10,nakano13,paula13,spehner13,spehner14} and methods 
	of estimating their upper and lower bounds are about five years old
	\cite{dakic10,miranowicz12,tufarelli13}. Note, that only the methods 
	for the calculation of the limits on the quantum discord are direct ones,
	allowing for calculation from the density matrix of the studied system --
	the calculation of precise values of discord still requires minimization over the set of all
	zero-discord states.
	
	\subsection{Separable states vs. zero-discord states \label{sec2a}}
	
	The class of separable states can be generally represented
	mathematically as the set of states, which can be written in the form,
	\begin{equation}
	\label{separablest}
	\rho_{\mathrm{AB}}^{\mathrm{sep}}=\sum_{\alpha}p_{\alpha}
	\rho_A^{\alpha}\otimes\rho_B^{\alpha}.
	\end{equation} 
	Here, the density matrices on the left side of the tensor product correspond to subsystem
	$A$ and the ones on the right side correspond to subsystem $B$.
	The only constraint is on the parameters of the decomposition
	$p_{\alpha}$, which have to be probabilities, $0<p_{\alpha}<1$
	and $\sum_{\alpha}p_{\alpha}=1$. Hence there are no constraints on the states
	$\rho_A^{\alpha}$ and $\rho_B^{\alpha}$, which do not have to
	be pure (although there does exist an equivalent definition using projectors) 
	or form an orthonormal basis for subsystem $A$ or $B$. 
	The lack of the orthonormality requirement
	is in fact the reason, why checking for entanglement between two subsystems
	is in general complicated.
	
	The class of zero-discord states can be represented in an analogous way \cite{modi14}.
	The only difference is that there is an additional constraint on states 
	of one or both subsystems.
	If the system state has zero discord with respect to subsystem $A$ ($B$),
	then there must exist a decomposition of the joint state of systems
	$A$ and $B$ such, that the density matrices $\rho_{A(B)}^{\alpha}$ in 
	eq.~(\ref{separablest}) can be written as projectors,
	\begin{equation}
	\rho_{A(B)}^{\alpha}=|a_{\alpha}\rangle\langle a_{\alpha}|,
	\end{equation}
	where
	$\{|a_{\alpha}\rangle\}$ forms
	an ortonormal set in the subspace of subsystem $A$ ($B$). 
	If both the zero-discord criteria for subsystems $A$ and $B$ are fulfilled
	then the state is completely zero-discordant (there is no discord with respect
	to either subsystem).
	Note, that the set of zero-discord states
	is obviously a subset of separable states
	regardless of whether it is discordant with respect to one or both subsystems.
	
	\subsection{The criteria for zero-discord states \label{sec3}}
	
	Contrarily to the case of entanglement, for which even the determination, if a
	mixed-state density matrix is entangled or not, is complicated for bipartite
	entanglement of larger systems (for which at least one is not a qubit or qutrit),
	the 
	determination, if the quantum discord is present in a system is fairly 
	straightforward even in case of two arbitrarily large systems
	\cite{dakic10,huang11}. 
	
	In the following we used the criterion of Ref.~\cite{huang11}, which is more suitable for the class of systems under
	study (both criteria allow to check, if a state is discordant with respect
	to one of the potentially correlated systems at a time). 
	The criterion introduced in the paper states that a bipartite state
	(where the parties are of arbitrary dimension $N$ and $M$) has zero quantum discord
	with respect to the system $M$,
	if and only if all blocks of its density matrix, after the bipartite
	$(NM)\times(NM)$ density matrix is partitioned into $N^2$ matrices
	of dimension $M\times M$
	(the particulars of the partition are described below), 
	are normal matrices and
	commute with each other.
	
	Here, the partition is performed starting from a bipartite density matrix
	\begin{equation}
	\hat{\sigma}=\sum_{kq}\sum_{nm}P_{kq}^{nm}|k\rangle \langle q|\otimes
	|n\rangle \langle m|,
	\end{equation}
	where the indices (and states labeled by them) $k,q$
	correspond to one of the subsystems (say, the one of dimension $N$), while the indices (and states)
	$n, m$ correspond to the other subsystem (of dimension $M$). Note, that the parameters 
	$P_{kq}^{nm}$ must fulfill a number of conditions for the matrix $\sigma$
	to be a density matrix, but we will not concern ourself with those here,
	since in the following a state obtained via
	a unitary evolution from an initial product state of two density matrices
	will be considered, which
	can obviously always be described by a density matrix).
	The partition of this density matrix into $N^2$ blocks
	of dimension $M\times M$ is done as follows
	\begin{equation}
	\label{partition}
	\hat{\sigma}_{kq}=\langle k|\hat{\sigma}| q\rangle,
	\end{equation}
	for all $k,q$.
	
	The criterion of normality means that for all $k,q$
	\begin{equation}
	[\hat{\sigma}_{kq},\hat{\sigma}^{\dagger}_{kq}]=0,
	\end{equation}
	while the commutation criterion means that for all $k,q$ and $k',q'$
	\begin{equation}
	[\hat{\sigma}_{kq},\hat{\sigma}_{k'q'}]=0.
	\end{equation}
	Both criteria are fulfilled, if and only if the state has zero discord
	with respect to subsystem $M$
	(meaning that the state of subsystem of size $M$ can be fully determined
	by local measurements performed on this subsystem and 
	classical communication alone
	without disturbing it).
	
	\section{The class of systems under study \label{sec4}}
	
	We study a class of systems consisting of a qubit and an environment which, when only the qubit is a system of interest, always lead to pure dephasing evolutions
	of the qubit (the occupations of the qubit remain unchanged).
	In general, the Hamiltonian of such a system can be written as
	\begin{eqnarray}
	\label{H}
	\hat{H}&=&
	\hat{H}_{\mathrm{Q}}+\hat{H}_{\mathrm{E}}+ |0\rangle\langle 0|\otimes{\hat{V_0}} +|1\rangle\langle 1|\otimes{\hat{V_1}} \label{eq:Hgen} \,\, .
	\end{eqnarray}
	The first term of the Hamiltonian describes the qubit and is given by $\hat{H}_{\mathrm{Q}}=\sum_{i=0,1}\varepsilon_{i}|i\rangle\langle i|$,  the second  describes the environment and is arbitrary, while the remaining terms describe the qubit-environment interaction with the qubit states written on the left side of each term and the environmental operators $\hat{V_0}$ and $\hat{V_1}$ are also arbitrary. Such a pure dephasing model of decoherence is not only theoretically easier to treat than a more general case of arbitrary qubit-environment coupling - it is also applicable to a very wide class of experimentally relevant qubit systems. For a vast majority of qubits, inlcuding almost all the solid-state based ones \cite{Szankowski_JPCM17}, but also trapped ions \cite{Monz_PRL11}, pure dephasing of superposition of $\ket{0}$ and $\ket{1}$ pointer states happens on time scales much shorter than relaxation between these states caused by energy exchange with the environment, and basically all the coherence loss can be modeled with the above Hamiltonian. Some specific examples of systems described by Hamiltonian (\ref{eq:Hgen}) include excitonic qubits coupled to phonons \cite{roszak06b,Krzywda_SR16,Salamon_PRA17,Paz_PRA17} (then $\hat{H}_{\mathrm{E}}$ is a Hamiltonian of free bosonic modes, $\hat{V}_{0}\! =\! 0$ as $\ket{0}$ corresponds to no exciton, and $\hat{V}_{1}$ is linear in creation and annihilation operators of phonons), qubits based on single spins in quantum dots interacting with spin baths at finite magnetic fields \cite{Cywinski_PRB09,Yang_RPP17} (then $\hat{V}_{0} \! =\! -\hat{V}_{1}$ so that the coupling can be written as $\propto \hat{V}\hat{\sigma}_{z}$ where $\hat{V}$ is the operator of the nuclear Overhauser field and $\hat{\sigma}_{z}$ pertains to spin-$1/2$ qubit), and spin qubits based on nitrogen-vacancy (NV) center in diamond coupled to a bath of carbon nuclei \cite{Zhao_PRB12} or electron spins \cite{deLange_Science10,Witzel_PRB12}, for which both $\hat{V}_{0} = -\hat{V}_{1}$ and $\hat{V}_{0} \! =\! 0$, $\hat{V}_{1}\! \neq \! 0$ can be realized \cite{Zhao_PRL11}.
	
	The full qubit-environment evolution operator $\hat{U}(t)\! =\! \exp(-i\hat{H}t)$ resulting from the Hamiltonian (\ref{eq:Hgen}) can be written as
	\begin{equation}
	\label{u}
	\hat{U}(t) = |0\rangle\langle 0|\otimes\hat{w}_0(t)+ |1\rangle\langle 1|\otimes \hat{w}_1(t) \,\, ,
	\end{equation}
	where we have defined the operators 
	\begin{equation}
	\label{w}
	\hat{w}_i(t)= \exp(-i\hat{H}_{i}t) \,\, ,
	\end{equation}
	with $i=0,1$, and $\hat{H}_{i} \! =\! \hat{H}_{\mathrm{E}} + \hat{V}_{i}$.
	
	We study the joint state of a qubit and an environment which are initially in a product state $\hat{\sigma}(0)=\hat{\rho}_{\mathrm{Q}}(0)\otimes\hat{R}(0)$ and evolve according to the operator (\ref{u}).
	The qubit is initially in a pure state $|\psi\rangle=\alpha|0\rangle+\beta|1\rangle$, with
	$|\alpha|^2+|\beta|^2=1$ and $\alpha,\beta\neq 0$ (a superposition is needed for dephasing to occur
	as well as entanglement and discord generation), 
	so the density matrix $\hat{\rho}_{\mathrm{Q}}(0)=|\psi\rangle\langle\psi |$.
	We impose no restrictions on the initial density matrix of the environment and write it in terms of its  eigenstates, $\hat{R}(0)=\sum_n c_n|n\rangle\langle n|$. The time-evolved qubit-environment density matrix
	takes the form
	\begin{equation}
	\begin{array}{l}
	\label{mac1}
	\hat{\sigma}(t)=\\
	\left(
	\begin{array}{cc}
	|\alpha|^2\sum_n c_n|n_0(t)\rangle\langle n_0(t)|&\alpha\beta^*\sum_n c_n|n_0(t)\rangle\langle n_!(t)|\\
	\alpha^*\beta\sum_n c_n|n_1(t)\rangle\langle n_0(t)|&|\beta|^2\sum_n c_n|n_1(t)\rangle\langle n_1(t)|
	\end{array}\right),
	\end{array}
	\end{equation}
	where the matrix is written in the basis of the eigenstates of the free qubit
	Hamiltonian, and $|n_i(t)\rangle=\hat{w}_i(t)|n\rangle$ with $\hat{w}_i(t)$ given by Eq.~(\ref{w}).
	
	Note, that the density matrix written as in eq.~(\ref{mac1}) is already decomposed
	into four blocks (with respect to the states of the
	qubit). Each element of the matrix (\ref{mac1}) written as it is in block form,
	is the type of block
	that allows to check for the presence of the discord with respect to the
	environment following the criteria
	of the previous section.
	
	\subsection{The zero-entanglement criterion for pure dephasing \label{sec5}}
	
	The problem of separability for the class of systems described in the previous
	section has been solved in Ref.~\cite{roszak15a}. A joint state of the qubit
	and its environment (\ref{mac1}) which is generated by the evolution operator
	given by eq.~(\ref{u}) (which itself comes from the Hamiltonian (\ref{eq:Hgen}))
	is separable at time $t$, if and only if 
	\beq
	[\hat{R}(0),\hat{w}(t)] \! = \!0 \, \, , \label{eq:main}
	\eeq
	where
	\begin{equation}
	\label{w}
	\hat{w}(t)= \exp(i\hat{H}_{0}t)\exp(-i\hat{H}_{1}t)= \hat{w}_0(t)\hat{w}_1(t)\,\, .
	\end{equation}
	This criterion can be equivalently stated as
	\begin{equation}
	\label{ent_env_inaczej}
	\hat{w}^{\dagger}_0(t)\hat{R}(0)\hat{w}_0(t)=
	\hat{w}^{\dagger}_1(t)\hat{R}(0)\hat{w}_1(t).
	\end{equation}
	In this form, the criterion is particularly easy to compare with the results
	of applying the zero-discord criteria to the system under study.
	
	\section{Equivalence of separability and zero-discord with respect
		to the environment \label{sec6}}
	
	Using the criterion introduced in Sec.~\ref{sec3} to check,
	if the
	system-environment 
	density matrix given by eq.~(\ref{mac1}) is discordant
	with respect to the environment at time $t$
	is uncomplicated. We have the density matrix of the qubit and
	environment written in such a way that it is already divided into the
	aforementioned blocks,
	meaning that each of the four matrices in the subspace of the environment 
	$\hat{\sigma}_{ij}(t)=\langle i|\hat{\sigma}(t)|j\rangle$
	is a separate block ($|i\rangle,|j\rangle=|0\rangle,|1\rangle$
	are qubit states).
	
	It is straightforward to show that $\hat{\sigma}_{00}(t)$ and 
	$\hat{\sigma}_{11}(t)$ are always normal, since $\hat{\sigma}_{00}(t)=\hat{\sigma}_{00}^{\dagger}(t)$ and 
	$\hat{\sigma}_{11}(t)=\hat{\sigma}_{11}^{\dagger}(t)$,
	so $[\hat{\sigma}_{ii}(t),\hat{\sigma}_{ii}^{\dagger}(t)]=0$.
	For the blocks corresponding to the diagonal elements of the qubit
	density matrix, $\hat{\sigma}_{01}(t)=\hat{\sigma}_{10}^{\dagger}(t)$,
	the part of the normality criterion for zero-discord states,
	is not so easy to check and in fact these matrices are not always normal.
	We will return to this in the next paragraph, since this normality criterion
	is equivalent to one of the commutation criteria.

	Before we continue, let us denote 
	\begin{equation}
	\hat{R}_{ij}(t)=\hat{w}_i(t)\hat{R}(0)\hat{w}_j^{\dagger}(t).
	\end{equation}
	The criterion that all $\hat{\sigma}_{ij}(t)$ must commute, obviously reduces to the criterion that all $\hat{R}_{ij}(t)$ must commute.
	For the density matrix (\ref{mac1}), this leads to the following 
	commutation conditions
	\begin{subequations}
		\label{nodiscord}
	\begin{eqnarray}
	\left[\hat{R}_{00}(t),\hat{R}_{11}(t)\right]&=&0,\\
	\left[\hat{R}_{00}(t),\hat{R}_{01}(t)\right]&=&
	\hat{R}_{01}(t)\left(\hat{R}_{11}(t)-\hat{R}_{00}(t)\right)=0,\\
	\left[\hat{R}_{00}(t),\hat{R}_{10}(t)\right]&=&
	\left(\hat{R}_{00}(t)-\hat{R}_{11}(t)\right)\hat{R}_{10}(t)=0,\\
	\left[\hat{R}_{11}(t),\hat{R}_{01}(t)\right]&=&
	\left(\hat{R}_{11}(t)-\hat{R}_{00}(t)\right)\hat{R}_{01}(t)=0,\\
	\left[\hat{R}_{11}(t),\hat{R}_{10}(t)\right]&=&
	\hat{R}_{10}(t)\left(\hat{R}_{00}(t)-\hat{R}_{11}(t)\right)=0,\\
	\label{comcrit}
	\left[\hat{R}_{01}(t),\hat{R}_{10}(t)\right]&=&\hat{R}_{00}(t)^2-\hat{R}_{11}(t)^2=0,
	\end{eqnarray}
	\end{subequations}
	since
	\begin{eqnarray}
	\nonumber
	\hat{R}_{00}(t)\hat{R}_{01}(t)&=&
	\hat{w}_0(t)\hat{R}(0)\hat{w}_0^{\dagger}(t)
	\hat{w}_0(t)\hat{R}(0)\hat{w}_1^{\dagger}(t)\\
	\nonumber
	&=&
	\hat{w}_0(t)\hat{R}(0)\hat{w}_1^{\dagger}(t)
	\hat{w}_1(t)\hat{R}(0)\hat{w}_1^{\dagger}(t)\\
	&=&
	\hat{R}_{01}(t)\hat{R}_{11}(t),
	\end{eqnarray}
	etc..
	Note, that the criteria of normality for matrices $\hat{\sigma}_{01}$ and
	$\hat{\sigma}_{10}$ are always equivalent to commutation criterion (\ref{comcrit}),
	since $\hat{R}_{01}^{\dagger}(t)=\hat{R}_{10}(t)$.
	Furthermore, 
	all of the conditions (\ref{nodiscord}) are always satisfied 
	when the state (\ref{mac1}) is separable, since then 
	$\hat{R}_{00}(t)-\hat{R}_{11}(t)=0$ as shown in eq.~(\ref{ent_env_inaczej}).
	
	This means, quite surprisingly, that the
	class of separable (zero-entanglement) qubit-environment states which can be obtained during
	a pure-dephasing evolution is equivalent to the class of states with zero discord
	with respect to the environment 
	for this type of evolution, since entangled states are always also
	discordant \cite{modi14}. Consequently, for this type of evolution
	there is little or no difference between entanglement and the
	environmental quantum discord.
	The discord may still display qualities resulting from points of indifferentiability, but entanglement
	evolution is unlikely to display its most characteristic feature, namely sudden death of
	entanglement, because for the class of systems under study not only the set
	of zero-discord states has zero volume, but also the set of separable states
	has zero volume (even ``one-sided'' zero-discord states possess the 
	zero-volume quality).
	
	\section{Separability and zero-discord with respect to the qubit \label{sec6a}}
	
	The relationship between separability and the lack of discord with respect
	to the qubit subspace is much more ambiguous. 
	Since local unitary operations cannot change the amount of discord
	in a system \cite{hassan13} and specifically
	no local operations on the environment 
	can change, whether a state is discordant with respect to the qubit or not,
	as is evident from the definition of ``one-sided'' zero-discord states in Sec.~{\ref{sec2a}},
	nor do they change the purity of the reduced
	density matrix of the qubit,
	let us work with the qubit-environment density matrix transformed by an unitary operation on the environment (as we did in  \cite{roszak15a}),
	namely
	\begin{equation}
	\label{mac2}
	\tilde{\sigma}(t)=\hat{w}_0^{\dagger}(t)\hat{\sigma}(t)\hat{w}_0(t).
	\end{equation}
	Since separability at time $t$ 
	indicates that there exists a (time-dependent) basis $\{|n'(t)\rangle\}$ in which
	both the initial density matrix of the environment $\hat{R}(0)$ and the operator
	$\hat{w}_0^{\dagger}(t)\hat{w}_1(t)$ are diagonal, they can be written as
	$\hat{R}(0)=\sum_n c_n'(t)|n'(t)\rangle\langle n'(t)|$ and
	$\hat{w}_0^{\dagger}(t)\hat{w}_1(t)=\sum_n \exp(-i\phi_n(t))|n'(t)\rangle\langle n'(t)|$.
	Consequently the density matrix (\ref{mac2}) can be written as
	\begin{equation}
	\label{mac3}
	\tilde{\sigma}(t)=\sum_n c_n'(t)
	\left(
	\begin{array}{cc}
	|\alpha|^2&\alpha\beta^*e^{i\phi_n(t)}\\
	\alpha^*\beta e^{-i\phi_n(t)}&|\beta|^2
	\end{array}\right)\otimes|n'(t)\rangle\langle n'(t)|.
	\end{equation}
	
	If the separable system density matrix corresponding to time $t$, 
	eq.~(\ref{mac3}), is partitioned into $N^2$
	(where $N$ is the dimension of the environment) $2\times 2$ matrices
	in terms of the eigenbasis $\{|n'(t)\rangle\}$,
	then the diagonal matrices (the ones corresponding to
	$|k\rangle=|q\rangle=|n'(t)\rangle$
	in eq.~(\ref{partition})) are of the form
	\begin{equation}
	\langle n'(t)|\tilde{\sigma}(t)|n'(t)\rangle=c_n'(t)\left(
	\begin{array}{cc}
	|\alpha|^2&\alpha\beta^*e^{-i\phi_n(t)}\\
	\alpha^*\beta e^{i\phi_n(t)}&|\beta|^2
	\end{array}\right),
	\end{equation}
	while the off-diagonal matrices are equal to zero.
	Hence, all of the matrices of this partition fulfill the normality requirement
	for the discord,
	while the commutation requirements are reduced to 
	\begin{widetext}
	\begin{equation}
		\begin{array}{l}
	\label{szeroki}
	\left[\langle n'(t)|\tilde{\sigma}(t)|n'(t)\rangle,
	\langle m'(t)|\tilde{\sigma}(t)|m'(t)\rangle\right]\\
	=
	c_n'(t)c_m'(t)\left(
	\begin{array}{cc}
	2i|\alpha|^2|\beta|^2\sin\left[\phi_n(t)-\phi_m(t)\right]
	&-\alpha\beta^*(|\alpha|^2-|\beta|^2)\left(e^{i\phi_n(t)}-e^{i\phi_m(t)}\right)\\
	\alpha\beta^*(|\alpha|^2-|\beta|^2)\left(e^{-i\phi_n(t)}-e^{-i\phi_m(t)}\right)&
	2i|\alpha|^2|\beta|^2\sin\left[\phi_n(t)-\phi_m(t)\right]
	\end{array}
	\right)=0
	\end{array}
	\end{equation}
	\end{widetext}
	for all $n$ and $m$.
	
	Firstly, let us identify the trivial solutions  that lead to no
	pure dephasing of the qubit (when no correlations
	of any type between a qubit and its environment are formed and the qubit-environment
	state remains a product). These include
	an initial state of the qubit which is not a superposition, i.e.~$\alpha=0$ or $\beta=0$,
	and the situation when for all $n$ and $m$ for which $c_n(t)'\neq 0$
	and $c_m'(t)\neq 0$, $\phi_n(t)=\phi_m(t)$ mod $2\pi$.
	Furthermore, the condition stemming from the off-diagonal elements of the matrices
	(\ref{szeroki}) that needs to be taken into account when $|\alpha|\neq|\beta|$,
	implies that for all $n$ and $m$ for which $c_n'(t)\neq 0$ and $c_m'(t)\neq 0$
	we must have $\exp(i\phi_n(t))=\exp(i\phi_m(t))$; if this
	condition is fulfilled, it is easy to see that the qubit does not undergo 
	pure dephasing (and only a phase shift between the elements of its superposition).
	Hence pure dephasing is always accompanied by discord generation with
	respect to the qubit, as long as the initial state of the qubit
	is not an equal superposition state. 
	
	The only case when the qubit can experience pure dephasing due to an interaction
	with the environment which is not accompanied by discord with respect to the 
	qubit state is, if it is initially in an equal superposition state, $|\alpha|=|\beta|=1/\sqrt{2}$.
	Then the set of commutation conditions for zero discord are reduced to 
	$\sin\left[\phi_n(t)-\phi_m(t)\right]=0$ for all $n$ and $m$.
	To see that such a situation is possible let us study the simplest example,
	with the dimension of the environment $N=2$. Imagine that at a certain time $t$,
	the exponential factors are $\exp(i\phi_{0}(t))=1$ and $\exp(i\phi_{1}(t))=-1$,
	respectively. The level of coherence of the qubit (the amplitude of the
	off-diagonal element of its reduced density matrix) is governed by the function
	$|c_0'(t)-c_1'(t)|$ and the qubit is fully coherent for $|c_0'(t)-c_1'(t)|=1$ and in a 
	completely mixed state for $|c_0'(t)-c_1'(t)|=0$. Obviously, regardless of the values of
	$c_0'(t)$ and $c_1'(t)$, there is no discord of any kind between the qubit and the
	environment, but only for $c_0'(t)=0$ or $c_1'(t)=0$ is the qubit fully coherent,
	while the coherence of the qubit depends only on the mixedness of the
	initial density matrix of the environment. 
	In the case of a single qubit environment it is not possible for 
	a pure dephasing evolution to
	have zero qubit discord for all times, because this would require 
	the initial environmental state to be pure, and for such a state the dephasing of the qubit is equivalent to creation of qubit-environment entanglement \cite{Zurek_RMP03}.

	In general, in order to have zero discord with respect to the qubit initialized in an equal superposition state which undergoes
	non-entangling evolution the following conditions have to be met at time $t$.
	For all $n$ and $m$ that correspond to nonzero coefficients $c_n'(t)$
	and $c_m'(t)$, the phase differences fulfill $\phi_n(t)-\phi_m(t)=p\pi$,
	where $p$ is an integer. For an environment with $N> 2$
	this implies a rigid condition on all phases corresponding to non-zero
	coefficients $c_n'(t)$, which must all have the form $\phi_n(t)=\phi_0(t)+q\pi$,
	again with  integer $q$.
	
	The situation is peculiar, since decoherence is almost always accompanied by
	a buildup of the quantum discord with respect to the qubit.
	The one (very notable)
	exception allows for discord-less dephased states, when the qubit state
	is of high symmetry and this symmetry is mirrored by the state of the environment.
	In this case any level of decoherence is possible,
	so it is not only a minor aberration, but a different type of decoherence
	process in terms of correlations with an environemnt. 
		
	\section{Enhancement of two-qubit entanglement under local decoherence} \label{sec6b}
	
	It has been recently shown \cite{Rodriguez_JPA08}
	that lack of the quantum discord
	with respect to one of the subsystems in an initial state implies complete
	positivity of the reduced dynamics of this subsystem. 
	Hence, in case of evolution of the type discussed here, in the case of zero Q-E entanglement being generated, and thus zero discord with respect to the environment being generated, the evolution of the environment
	can be described by CP maps not only from the studied initial
	product state, but also setting the initial time to any time $t$.
	As we have shown in the previous section this is not usually the case
	for the evolution of the qubit, that can certainly be described by CP maps
	from the initial product state,
	but nothing can be said for most evolved states. Note that the full understanding of conditions for initial system-environment correlations that make the subsequent evolution of the reduced state of the system completely positive is a subject of ongoing research \cite{Buscemi_PRL14,Vacchini_SR16,Dominy_framework_CP_QIP16,Dominy_beyond_CP_QIP16}.
	
	The formalism presented in Ref.~\cite{roszak15a} in general cannot be used to study
	two-qubit states undergoing pure dephasing due to an interaction 
	with an environment, but it turns out that it is
	viable, if the initial qubit state is a superposition of only two states that are product states in the pointer states bases of the qubits (i.e.~the bases singled out by the pure dephasing couplings to the respective environments).
	Such a state is ``operationally'' a two-level system,
	since during pure dephasing it will never leave the subspace of these two states, and the two-qubit entanglement is then simply proportional to the modulus of the single nonzero coherence present in the reduced density matrix of the qubits
	\cite{Yu_QIC07,mazurek14a,Szankowski_QIP15,Bragar_PRB15}.
	This allows for the study of the generation of entanglement between any initial
	two-qubit Bell state and its environment, while the state undergoes pure dephasing
	and its entanglement is diminished.
	 
	Furthermore, if the qubits are initialized in a pure state,
	each qubit interacts with a separate environment, the initial state of these environments is a product  of density operators of the two environments, and there is no interaction between the qubits and between the environments, then the evolution
	of each qubit is local, so the qubits evolve under local operations and classical
	communication (LOCC). In the case of a product initial state 
	of the two-qubit state and the environments, It is known that such evolution
	cannot lead to enhancement of entanglement between the initial
	qubit state and any qubit state at time $t$ -- this is follows from definition of entanglement as a quantity that cannot be increased by LOCC \cite{Plenio_QIC07,horodecki09}. 
	
	However, it has to be stressed that while the above-listed conditions for LOCC dynamics are too strong (they are definitely sufficient, but not all of them are necessary) breaking of any of them could make the evolution become nonlocal. Full understanding of the necessary conditions is hampered by the fact that LOCC transformations are notoriously difficult to characterize (see \cite{Plenio_QIC07,Aolita_RPP15} and references therein). It is easier to consider a larger set (containing LOCC within it) of so-called separable maps that are CP and for which all the Kraus operators defining them can be written as products of operators acting on relevant local subsystems. This however comes at a price: there exist separable maps that can increase entanglement of states \cite{Chitambar_PRL09} that are neither ``generically'' separable nor pure \cite{Gheorghiu_PRA08}, e.g.~certain states that were obtained from pure entangled states by subjecting them to decoherence.
	
	We have reminded the reader about those known results in order to stress the fact that the mathematical conditions for an evolution not to increase entanglement are highly nontrivial, and consequently simple intuitions about what kind of evolution is ``local'' (and thus cannot increase entanglement) often prove wrong.
	It is known that local operations can enhance entanglement with respect to an initial state having correlations between the entangled system and the environment (for an example see \cite{DArrigo_AP14}), if these correlations make the subsequent dynamics not completely positive, and thus not belonging to LOCC, as LOCC is a subset of separable CP operations.
	As shown recently, nonzero discord between the qubits and their environment can be \cite{Rodriguez_JPA08} (but {\it does not have to be}, see \cite{Dominy_framework_CP_QIP16}) a correlation that leads to subsequent non-CP dynamics. Let us see then if the two-qubit generalization of the previously obtained results about system-environment discord generated during decoherence, can shed some light on the behavior of two-qubit entanglement dynamics.

	To this end let us study the initial two-qubit Bell state 
	$|\psi\rangle=1/\sqrt{2}(|00\rangle+|11\rangle)$. The choice of the Bell state
	is arbitrary. We assume (for simplicity) that only one of the qubits interacts
	with an environment and that this environment is a qubit itself in some 
	initial state $R(0)=c_0|0\rangle\langle 0|+c_1|1\rangle\langle 1|$.
	The product initial state of the whole state system is therefore 
	$\hat{\sigma}(0)=|\psi\rangle\langle \psi |\otimes R(0)$.
	The Hamiltonian of this system is
	\begin{equation}
	\hat{H}=\varepsilon_A|1\rangle_{AA}\langle 1|
	+\varepsilon_B|1\rangle_{BB}\langle 1|+
	|1\rangle_{AA}\langle 1|\otimes\hat{V}_A+\hat{H}_E,
	\end{equation}
	where the indices $A$ and $B$ differentiate between the qubits,
	$\hat{V}_A$ is an operator acting in the subspace of the environment,
	while $\hat{H}_E$ is the free Hamiltonian of the environment.
	The interaction of qubit $A$ with the environment has been asymmetrized
	for convenience, since the aim of this section is to show an exemplary evolution
	of a certain type
	and not to quantify all possible evolutions of this type.
	
	Although this Hamiltonian is of larger dimensionally in terms of the qubits
	than the Hamiltonian of eq.~(\ref{H}), the resulting evolution is equivalent
	to the evolution discussed in Sec.~\ref{sec4}, if the assumptions introduced
	in the previous paragraph are taken into account.
	Although the evolution operator is different and is equal to
	\begin{eqnarray}
	\hat{U}(t)&=&|00\rangle\langle 00|\otimes\unit+|01\rangle\langle 01|\otimes\unit\\
	\nonumber
	&&+|10\rangle\langle 10|\otimes\hat{w}(t)
	+|11\rangle\langle 11|\otimes\hat{w}(t),
	\end{eqnarray}
	where $\hat{w}(t)=\exp\left(-i(\hat{H}_E+\hat{V})t
	\right)$, the density matrix of the whole system evolves according to
	\begin{equation}
	\hat{\sigma}(t)=
	\frac{1}{2}\left(
	\begin{array}{cccc}
	\hat{R}(0)&0&0&\hat{R}(0)\hat{w}(t)\\
	0&0&0&0\\
	0&0&0&0\\
	\hat{w}^{\dagger}(t)\hat{R}(0)&0&0&\hat{w}^{\dagger}(t)\hat{R}(0)\hat{w}(t)
	\end{array}
	\right),
	\end{equation}
	which is the same as in case of a single qubit interacting with the type
	of environment under study.
	
	Let us now additionally assume that the evolution is non-entangling
	(as always in this paper), so it is possible to write $\hat{R}(0)$
	and $\hat{w}(t)$ in the common eigenbasis $\{|n'(t)\rangle\}$ at every time $t$.
	Hence, the state $\hat{\sigma}(t)$ can be written in the form
	\begin{equation}
	\hat{\sigma}(t)=\sum_{n=0}^{1}c_n'(t)|\psi_n(t)\rangle\langle\psi_n(t)|\otimes
	|n'(t)\rangle\langle n'(t)|,
	\end{equation}
	where $|\psi_n(t)\rangle =1/\sqrt{2}\left(|00\rangle+e^{i\phi_n(t)}|11\rangle
	\right)$ and the factors $e^{i\phi_n(t)}$ are eigenvalues of $\hat{w}(t)$ 
	corresponding to the eigenstates $|n'(t)\rangle$ appropriately.
	It is straightforward to quantify inter-qubit entanglement 
	in this state (after tracing-out the degrees of freedom of the environment)
	and the entanglement measure
	concurrence \cite{wootters98} of such a state is equal to
	$C_Q=|c_0'(t)e^{i\phi_0(t)}+c_1'(t)e^{i\phi_1(t)}|$.

	\begin{figure}[th]
		\begin{center}
			\unitlength 1mm
			\begin{picture}(75,55)(5,5)
			\put(0,0){\resizebox{85mm}{!}{\includegraphics{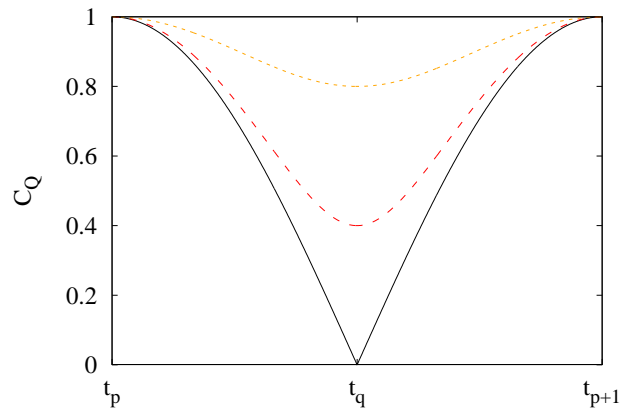}}}
			\end{picture}
		\end{center}
		\caption{\label{fig} Full cycle (evolution from $t_p$ to $t_{p+1}$) of inter-qubit entanglement when only one of the
			qubits interacts with an $N=2$ environment for $c_0=0.5$ (black solid line)
			$c_0=0.7$ (red dashed line), and $c_0=0.9$ (orange dotted line).}
	\end{figure}
	
	In the simplest case, the basis $\{|n'(t)\rangle\}$ is time independent
	and is the same as the eigenbasis of the initial state of the environment,
	$\{|0\rangle,|1\rangle\}$, so $c_0'(t)=c_0$ and $c_1'(t)=c_1$, while
	the time-dependence of the phase factors reduces to $\phi_n(t)=\varphi_nt$.
	Here, zero discord with respect to the qubit system is obtained only in two situations;
	firstly, at times $t_p$ when $|\psi_0(t_p)\rangle =|\psi_1(t_p)\rangle$,
	so $e^{i\varphi_0t_p}=e^{i\varphi_1t_p}$
	and the qubit is in a pure, maximally entangled state, and secondly,
	at times $t_q$ when $\langle\psi_0(t_q)|\psi_1(t_q)\rangle=0$, so
	$e^{i\varphi_0t_q}=-e^{i\varphi_1t_q}$
	and the qubit decoherence
	is maximal while inter-qubit entanglement is minimal. Times $t_p$ and $t_q$ appear interchangeably, since $t_p=2\pi p/|\varphi_1-\varphi_0|$
	and $t_q=2\pi (q+1)/|\varphi_1-\varphi_0|$, with $p,q=0,1,2...$
	and the evolution of inter-qubit entanglement is periodically repeated
	every $2\pi/|\varphi_1-\varphi_0|$.
	The evolution of such entanglement from a certain time $t_p$ to time $t_{p+1}$
	(capturing a full cycle of entanglement evolution)
	is shown in Fig.~(\ref{fig}) for three different initial states of the environment
	(the time $t_q$ which appears midcycle is marked on the figure).
	
	If we now choose a certain time $\tau =t_q$ as a new initial time, the evolution
	of the new initial
	state $\hat{\sigma}(\tau)$, which has minimal entanglement (zero entanglement
	for $c_0=c_1=1/2$) to any later state $t+\tau$ can be descibed using
	CP maps, since the state is zero-discordant with respect to the two qubits
	\cite{Rodriguez_JPA08}. 
	Note that all later states have greater or equal inter-qubit entanglement
	than the new initial state, so the evolution discussed in this section, which
	{\it could easily be mistakenly believed to be local, as it occurs due to interaction with an environment of only one of the qubits}, is also entangling, and thus it does not belong to LOCC. Apparently, the initial nonlocality (entanglement) of the qubits' state makes the total system state at time $t_{q}$ incompatible with subsequently local dynamics, although, as we have shown, this incompatibility does not follow from nonzero discord.
	Finally, let us note that we have so far failed at finding a separable representation of the CP evolution of the qubits from time $t_{q}$ onwards, but have not proven that such a representation is impossible. It remains then an open question if the example evolution described above is CP and separable, but not LOCC, or if it is simply CP but nonseparable. 
		Other examples of bipartite system and environment evolutions which lead
		to similar effects of entanglement increase (but not during pure dephasing)
		can be found in Ref.~\cite{krisnanda17,sargolzahi17}.
		
		It is interesting to further note that at all times the entanglement
		between one of the qubits and a subsystem containing the other qubit and
		the environment is maximal. This is easily checked by calculating the 
		entanglement measure Negativity \cite{plenio05b}. This fact is in full agreement with a theorem
		proven in Ref.~\cite{krisnanda17}, which for the scenario studied here
		reduces to the following (all conditions for the applicability of the theorem
		are fulfilled): If there is no discord between the two-qubit subsystem and the environment with respect to the environment, then entanglement between
		a single qubit and the rest of the system cannot increase with time. 
		Since the evolution of the whole system periodically
		returns to its initial state, entanglement of such partitions
		cannot change and remains constant.
	
	\section{Conclusion \label{sec7}}
	We have studied pure dephasing evolution of a qubit, intialized in a pure state, interacting with an environment
	and found that, if no qubit-environment entanglement is generated at a given time,  then automatically no 
	qubit-environment quantum discord with respect to the environment is generated.
	Hence, the set of separable states which can be obtained due to an evolution
	described by the class of Hamiltonians studied is zero-volume, and behaviors
	such as sudden death of qubit-environment entanglement are unlikely. Furthermore, the evolution
	of the environment alone between two arbitrary times may be described using 
	completely positive maps, as follows from connection between zero discord with respect to one subsystem and complete positivity of subsequent evolution of the reduced state of this system \cite{Rodriguez_JPA08}.
	
	We have also looked at the qubit-environment quantum discord with respect to 
	the qubit and it turns out that the situation is very different here.
	For times at which the qubit and the environment are separable, this type of discord
	is usually still present in the system. It is only possible for such discord
	to be zero, if the qubit is initially in an equal superposition state.
	Then the discord can vanish at certain times when the relative phases of the environmental states evolving due to the interaction with the qubit
	specifically align -- something that can be expected to happen only for rather small environments. Furthermore, an evolution for which this type of quantum correlations would {\it never} appear is impossible if we ignore the trivial cases of evolutions not leading to any decoherence.
	
	Lastly, we were able to compare an exemplary evolution of two-qubit entanglement
	under the influence of a {\it local} interaction with an environment,
	with qubit-environment discord generation. We have shown that 
	such an evolution which displays zero-discordant points in time with respect to the
	two qubit subsystem is possible (but not common). This means that 
	the evolution starting from one of the zero-discord times
	can be described using completely positive maps. Since at such points the qubits are either
	fully entangled (as they return at these times to their initial state) or have minimal entanglement
	possible, the evolution starting from times corresponding to the former case
	are trivial, but those corresponding to the latter situation, lead to 
	enhancement of inter-qubit entanglement due to a local interaction whilst
	the evolution can be described by completely positive maps.
	This interesting observation illustrates how hard it is to identify a LOCC evolution when one abandons the common assumption of complete lack of correlations between the initial states of the entangled system and the environment.
	
	\section*{Acknowledgments}
	This work was funded from the Reseach Project number DEC-2012/07/B/ST3/03616 financed by the National Science Centre (NCN). We would like to thank the organizers of FoKA 2017 workshop. Interesting discussions with Tomek Paterek are gratefully acknowledged.


\begin{thebibliography}{67}%
	\makeatletter
	\providecommand \@ifxundefined [1]{%
		\@ifx{#1\undefined}
	}%
	\providecommand \@ifnum [1]{%
		\ifnum #1\expandafter \@firstoftwo
		\else \expandafter \@secondoftwo
		\fi
	}%
	\providecommand \@ifx [1]{%
		\ifx #1\expandafter \@firstoftwo
		\else \expandafter \@secondoftwo
		\fi
	}%
	\providecommand \natexlab [1]{#1}%
	\providecommand \enquote  [1]{``#1''}%
	\providecommand \bibnamefont  [1]{#1}%
	\providecommand \bibfnamefont [1]{#1}%
	\providecommand \citenamefont [1]{#1}%
	\providecommand \href@noop [0]{\@secondoftwo}%
	\providecommand \href [0]{\begingroup \@sanitize@url \@href}%
	\providecommand \@href[1]{\@@startlink{#1}\@@href}%
	\providecommand \@@href[1]{\endgroup#1\@@endlink}%
	\providecommand \@sanitize@url [0]{\catcode `\\12\catcode `\$12\catcode
		`\&12\catcode `\#12\catcode `\^12\catcode `\_12\catcode `\%12\relax}%
	\providecommand \@@startlink[1]{}%
	\providecommand \@@endlink[0]{}%
	\providecommand \url  [0]{\begingroup\@sanitize@url \@url }%
	\providecommand \@url [1]{\endgroup\@href {#1}{\urlprefix }}%
	\providecommand \urlprefix  [0]{URL }%
	\providecommand \Eprint [0]{\href }%
	\providecommand \doibase [0]{http://dx.doi.org/}%
	\providecommand \selectlanguage [0]{\@gobble}%
	\providecommand \bibinfo  [0]{\@secondoftwo}%
	\providecommand \bibfield  [0]{\@secondoftwo}%
	\providecommand \translation [1]{[#1]}%
	\providecommand \BibitemOpen [0]{}%
	\providecommand \bibitemStop [0]{}%
	\providecommand \bibitemNoStop [0]{.\EOS\space}%
	\providecommand \EOS [0]{\spacefactor3000\relax}%
	\providecommand \BibitemShut  [1]{\csname bibitem#1\endcsname}%
	\let\auto@bib@innerbib\@empty
	%</preamble>
	\bibitem [{\citenamefont {{Deutsch}}\ and\ \citenamefont
		{{Jozsa}}(1992)}]{deutsch92}%
	\BibitemOpen
	\bibfield  {author} {\bibinfo {author} {\bibfnamefont {D.}~\bibnamefont
			{{Deutsch}}}\ and\ \bibinfo {author} {\bibfnamefont {R.}~\bibnamefont
			{{Jozsa}}},\ }\bibfield  {title} {\enquote {\bibinfo {title} {{Rapid Solution
					of Problems by Quantum Computation}},}\ }\href {\doibase
		10.1098/rspa.1992.0167} {\bibfield  {journal} {\bibinfo  {journal} {Royal
				Society of London Proceedings Series A}\ }\textbf {\bibinfo {volume} {439}},\
		\bibinfo {pages} {553--558} (\bibinfo {year} {1992})}\BibitemShut {NoStop}%
	\bibitem [{\citenamefont {Ekert}\ and\ \citenamefont
		{Jozsa}(1996)}]{Ekert_RMP96}%
	\BibitemOpen
	\bibfield  {author} {\bibinfo {author} {\bibfnamefont {Artur}\ \bibnamefont
			{Ekert}}\ and\ \bibinfo {author} {\bibfnamefont {Richard}\ \bibnamefont
			{Jozsa}},\ }\bibfield  {title} {\enquote {\bibinfo {title} {Quantum
				computation and shor's factoring algorithm},}\ }\href {\doibase
		10.1103/RevModPhys.68.733} {\bibfield  {journal} {\bibinfo  {journal} {Rev.
				Mod. Phys.}\ }\textbf {\bibinfo {volume} {68}},\ \bibinfo {pages} {733}
		(\bibinfo {year} {1996})}\BibitemShut {NoStop}%
	\bibitem [{\citenamefont {Bennett}\ \emph {et~al.}(1993)\citenamefont
		{Bennett}, \citenamefont {Brassard}, \citenamefont {Cr\'epeau}, \citenamefont
		{Jozsa}, \citenamefont {Peres},\ and\ \citenamefont
		{Wootters}}]{Bennett_PRL93}%
	\BibitemOpen
	\bibfield  {author} {\bibinfo {author} {\bibfnamefont {Charles~H.}\
			\bibnamefont {Bennett}}, \bibinfo {author} {\bibfnamefont {Gilles}\
			\bibnamefont {Brassard}}, \bibinfo {author} {\bibfnamefont {Claude}\
			\bibnamefont {Cr\'epeau}}, \bibinfo {author} {\bibfnamefont {Richard}\
			\bibnamefont {Jozsa}}, \bibinfo {author} {\bibfnamefont {Asher}\ \bibnamefont
			{Peres}}, \ and\ \bibinfo {author} {\bibfnamefont {William~K.}\ \bibnamefont
			{Wootters}},\ }\bibfield  {title} {\enquote {\bibinfo {title} {Teleporting an
				unknown quantum state via dual classical and einstein-podolsky-rosen
				channels},}\ }\href {\doibase 10.1103/PhysRevLett.70.1895} {\bibfield
		{journal} {\bibinfo  {journal} {Phys.\ Rev.\ Lett.}\ }\textbf {\bibinfo
			{volume} {70}},\ \bibinfo {pages} {1895--1899} (\bibinfo {year}
		{1993})}\BibitemShut {NoStop}%
	\bibitem [{\citenamefont {Popescu}(1994)}]{Popescu_PRL94}%
	\BibitemOpen
	\bibfield  {author} {\bibinfo {author} {\bibfnamefont {Sandu}\ \bibnamefont
			{Popescu}},\ }\bibfield  {title} {\enquote {\bibinfo {title} {Bell's
				inequalities versus teleportation: What is nonlocality?}}\ }\href {\doibase
		10.1103/PhysRevLett.72.797} {\bibfield  {journal} {\bibinfo  {journal} {Phys.
				Rev. Lett.}\ }\textbf {\bibinfo {volume} {72}},\ \bibinfo {pages} {797}
		(\bibinfo {year} {1994})}\BibitemShut {NoStop}%
	\bibitem [{\citenamefont {Werner}(1989)}]{werner89}%
	\BibitemOpen
	\bibfield  {author} {\bibinfo {author} {\bibfnamefont {Reinhard~F.}\
			\bibnamefont {Werner}},\ }\bibfield  {title} {\enquote {\bibinfo {title}
			{Quantum states with einstein-podolsky-rosen correlations admitting a
				hidden-variable model},}\ }\href {\doibase 10.1103/PhysRevA.40.4277}
	{\bibfield  {journal} {\bibinfo  {journal} {Phys. Rev. A}\ }\textbf {\bibinfo
			{volume} {40}},\ \bibinfo {pages} {4277--4281} (\bibinfo {year}
		{1989})}\BibitemShut {NoStop}%
	\bibitem [{\citenamefont {Plenio}\ and\ \citenamefont
		{Virmani}(2007)}]{Plenio_QIC07}%
	\BibitemOpen
	\bibfield  {author} {\bibinfo {author} {\bibfnamefont {Martin~B.}\
			\bibnamefont {Plenio}}\ and\ \bibinfo {author} {\bibfnamefont {Shashank}\
			\bibnamefont {Virmani}},\ }\bibfield  {title} {\enquote {\bibinfo {title} {An
				introduction to entanglement measures},}\ }\href@noop {} {\bibfield
		{journal} {\bibinfo  {journal} {Quant.~Info.~Comput.}\ }\textbf {\bibinfo
			{volume} {7}},\ \bibinfo {pages} {1} (\bibinfo {year} {2007})}\BibitemShut
	{NoStop}%
	\bibitem [{\citenamefont {Horodecki}\ \emph {et~al.}(2009)\citenamefont
		{Horodecki}, \citenamefont {Horodecki}, \citenamefont {Horodecki},\ and\
		\citenamefont {Horodecki}}]{horodecki09}%
	\BibitemOpen
	\bibfield  {author} {\bibinfo {author} {\bibfnamefont {Ryszard}\ \bibnamefont
			{Horodecki}}, \bibinfo {author} {\bibfnamefont {Pawe\l{}}\ \bibnamefont
			{Horodecki}}, \bibinfo {author} {\bibfnamefont {Micha\l{}}\ \bibnamefont
			{Horodecki}}, \ and\ \bibinfo {author} {\bibfnamefont {Karol}\ \bibnamefont
			{Horodecki}},\ }\bibfield  {title} {\enquote {\bibinfo {title} {Quantum
				entanglement},}\ }\href {\doibase 10.1103/RevModPhys.81.865} {\bibfield
		{journal} {\bibinfo  {journal} {Rev. Mod. Phys.}\ }\textbf {\bibinfo {volume}
			{81}},\ \bibinfo {pages} {865--942} (\bibinfo {year} {2009})}\BibitemShut
	{NoStop}%
	\bibitem [{\citenamefont {{Ollivier}}\ and\ \citenamefont
		{{Zurek}}(2001)}]{ollivier01}%
	\BibitemOpen
	\bibfield  {author} {\bibinfo {author} {\bibfnamefont {H.}~\bibnamefont
			{{Ollivier}}}\ and\ \bibinfo {author} {\bibfnamefont {W.~H.}\ \bibnamefont
			{{Zurek}}},\ }\bibfield  {title} {\enquote {\bibinfo {title} {{Quantum
					Discord: A Measure of the Quantumness of Correlations}},}\ }\href {\doibase
		10.1103/PhysRevLett.88.017901} {\bibfield  {journal} {\bibinfo  {journal}
			{Phys. Rev. Lett.}\ }\textbf {\bibinfo {volume} {88}},\ \bibinfo {eid}
		{017901} (\bibinfo {year} {2001})},\ \Eprint
	{http://arxiv.org/abs/quant-ph/0105072} {quant-ph/0105072} \BibitemShut
	{NoStop}%
	\bibitem [{\citenamefont {{Henderson}}\ and\ \citenamefont
		{{Vedral}}(2001)}]{henderson01}%
	\BibitemOpen
	\bibfield  {author} {\bibinfo {author} {\bibfnamefont {L.}~\bibnamefont
			{{Henderson}}}\ and\ \bibinfo {author} {\bibfnamefont {V.}~\bibnamefont
			{{Vedral}}},\ }\bibfield  {title} {\enquote {\bibinfo {title} {{Classical,
					quantum and total correlations}},}\ }\href {\doibase
		10.1088/0305-4470/34/35/315} {\bibfield  {journal} {\bibinfo  {journal}
			{Journal of Physics A Mathematical General}\ }\textbf {\bibinfo {volume}
			{34}},\ \bibinfo {pages} {6899--6905} (\bibinfo {year} {2001})},\ \Eprint
	{http://arxiv.org/abs/arXiv:quant-ph/0105028} {arXiv:quant-ph/0105028}
	\BibitemShut {NoStop}%
	\bibitem [{\citenamefont {Modi}(2014)}]{modi14}%
	\BibitemOpen
	\bibfield  {author} {\bibinfo {author} {\bibfnamefont {Kavan}\ \bibnamefont
			{Modi}},\ }\bibfield  {title} {\enquote {\bibinfo {title} {A pedagogical
				overview of quantum discord},}\ }\href {\doibase 10.1142/S123016121440006X}
	{\bibfield  {journal} {\bibinfo  {journal} {Open Syst. Inf. Dyn.}\ }\textbf
		{\bibinfo {volume} {21}},\ \bibinfo {pages} {1440006} (\bibinfo {year}
		{2014})}\BibitemShut {NoStop}%
	\bibitem [{\citenamefont {Ferraro}\ \emph {et~al.}(2010)\citenamefont
		{Ferraro}, \citenamefont {Aolita}, \citenamefont {Cavalcanti}, \citenamefont
		{Cucchietti},\ and\ \citenamefont {Ac{\'i}n}}]{ferraro10}%
	\BibitemOpen
	\bibfield  {author} {\bibinfo {author} {\bibfnamefont {A.}~\bibnamefont
			{Ferraro}}, \bibinfo {author} {\bibfnamefont {L.}~\bibnamefont {Aolita}},
		\bibinfo {author} {\bibfnamefont {D.}~\bibnamefont {Cavalcanti}}, \bibinfo
		{author} {\bibfnamefont {F.~M.}\ \bibnamefont {Cucchietti}}, \ and\ \bibinfo
		{author} {\bibfnamefont {A.}~\bibnamefont {Ac{\'i}n}},\ }\bibfield  {title}
	{\enquote {\bibinfo {title} {Almost all quantum states have nonclassical
				correlations},}\ }\href@noop {} {\bibfield  {journal} {\bibinfo  {journal}
			{Phys. Rev. A}\ }\textbf {\bibinfo {volume} {81}},\ \bibinfo {pages} {052318}
		(\bibinfo {year} {2010})}\BibitemShut {NoStop}%
	\bibitem [{\citenamefont {Rajagopal}\ and\ \citenamefont
		{Rendell}(2001)}]{rajagopal01}%
	\BibitemOpen
	\bibfield  {author} {\bibinfo {author} {\bibfnamefont {A.~K.}\ \bibnamefont
			{Rajagopal}}\ and\ \bibinfo {author} {\bibfnamefont {R.~W.}\ \bibnamefont
			{Rendell}},\ }\bibfield  {title} {\enquote {\bibinfo {title} {Decoherence,
				correlation, and entanglement in a pair of coupled quantum dissipative
				oscillators},}\ }\href {\doibase 10.1103/PhysRevA.63.022116} {\bibfield
		{journal} {\bibinfo  {journal} {Phys. Rev. A}\ }\textbf {\bibinfo {volume}
			{63}},\ \bibinfo {pages} {022116} (\bibinfo {year} {2001})}\BibitemShut
	{NoStop}%
	\bibitem [{\citenamefont {\ifmmode~\dot{Z}\else \.{Z}\fi{}yczkowski}\ \emph
		{et~al.}(2001)\citenamefont {\ifmmode~\dot{Z}\else \.{Z}\fi{}yczkowski},
		\citenamefont {Horodecki}, \citenamefont {Horodecki},\ and\ \citenamefont
		{Horodecki}}]{zyczkowski01}%
	\BibitemOpen
	\bibfield  {author} {\bibinfo {author} {\bibfnamefont {Karol}\ \bibnamefont
			{\ifmmode~\dot{Z}\else \.{Z}\fi{}yczkowski}}, \bibinfo {author}
		{\bibfnamefont {Pawe\l{}}\ \bibnamefont {Horodecki}}, \bibinfo {author}
		{\bibfnamefont {Micha\l{}}\ \bibnamefont {Horodecki}}, \ and\ \bibinfo
		{author} {\bibfnamefont {Ryszard}\ \bibnamefont {Horodecki}},\ }\bibfield
	{title} {\enquote {\bibinfo {title} {Dynamics of quantum entanglement},}\
	}\href {\doibase 10.1103/PhysRevA.65.012101} {\bibfield  {journal} {\bibinfo
		{journal} {Phys. Rev. A}\ }\textbf {\bibinfo {volume} {65}},\ \bibinfo
	{pages} {012101} (\bibinfo {year} {2001})}\BibitemShut {NoStop}%
\bibitem [{\citenamefont {Yu}\ and\ \citenamefont {Eberly}(2006)}]{yu06}%
\BibitemOpen
\bibfield  {author} {\bibinfo {author} {\bibfnamefont {T.}~\bibnamefont
		{Yu}}\ and\ \bibinfo {author} {\bibfnamefont {J.H.}\ \bibnamefont {Eberly}},\
}\bibfield  {title} {\enquote {\bibinfo {title} {Sudden death of
		entanglement: Classical noise effects},}\ }\href {\doibase
https://doi.org/10.1016/j.optcom.2006.01.061} {\bibfield  {journal} {\bibinfo
	{journal} {Optics Communications}\ }\textbf {\bibinfo {volume} {264}},\
\bibinfo {pages} {393 -- 397} (\bibinfo {year} {2006})},\ \bibinfo {note}
{quantum Control of Light and MatterIn honor of the 70th birthday of Bruce
	Shore}\BibitemShut {NoStop}%
\bibitem [{\citenamefont {Daki{\'c}}\ \emph {et~al.}(2010)\citenamefont
	{Daki{\'c}}, \citenamefont {Vedral},\ and\ \citenamefont
	{Brukner}}]{dakic10}%
\BibitemOpen
\bibfield  {author} {\bibinfo {author} {\bibfnamefont {B.}~\bibnamefont
		{Daki{\'c}}}, \bibinfo {author} {\bibfnamefont {V.}~\bibnamefont {Vedral}}, \
	and\ \bibinfo {author} {\bibfnamefont {{\v C}.}~\bibnamefont {Brukner}},\
}\bibfield  {title} {\enquote {\bibinfo {title} {Neccesary and sufficient
		condition for nonzero quantum discord},}\ }\href@noop {} {\bibfield
{journal} {\bibinfo  {journal} {Phys. Rev. Lett.}\ }\textbf {\bibinfo
	{volume} {105}},\ \bibinfo {pages} {190502} (\bibinfo {year}
{2010})}\BibitemShut {NoStop}%
\bibitem [{\citenamefont {Roszak}\ and\ \citenamefont
	{Cywi{\'n}ski}(2015)}]{roszak15b}%
\BibitemOpen
\bibfield  {author} {\bibinfo {author} {\bibfnamefont {K.}~\bibnamefont
		{Roszak}}\ and\ \bibinfo {author} {\bibfnamefont {{\L}.}~\bibnamefont
		{Cywi{\'n}ski}},\ }\bibfield  {title} {\enquote {\bibinfo {title} {The
			relation between the quantum discord and quantum teleportation: The physical
			interpretation of the transition point between different quantum discord
			decay regimes},}\ }\href {http://stacks.iop.org/0295-5075/112/i=1/a=10002}
{\bibfield  {journal} {\bibinfo  {journal} {EPL (Europhysics Letters)}\
	}\textbf {\bibinfo {volume} {112}},\ \bibinfo {pages} {10002} (\bibinfo
	{year} {2015})}\BibitemShut {NoStop}%
\bibitem [{\citenamefont {Mazzola}\ \emph {et~al.}(2010)\citenamefont
	{Mazzola}, \citenamefont {Piilo},\ and\ \citenamefont
	{Maniscalco}}]{mazzola10}%
\BibitemOpen
\bibfield  {author} {\bibinfo {author} {\bibfnamefont {L.}~\bibnamefont
		{Mazzola}}, \bibinfo {author} {\bibfnamefont {J.}~\bibnamefont {Piilo}}, \
	and\ \bibinfo {author} {\bibfnamefont {S.}~\bibnamefont {Maniscalco}},\
}\bibfield  {title} {\enquote {\bibinfo {title} {Sudden transition between
		classical and quantum decoherence},}\ }\href {\doibase
10.1103/PhysRevLett.104.200401} {\bibfield  {journal} {\bibinfo  {journal}
	{Phys. Rev. Lett.}\ }\textbf {\bibinfo {volume} {104}},\ \bibinfo {pages}
{200401} (\bibinfo {year} {2010})}\BibitemShut {NoStop}%
\bibitem [{\citenamefont {Roszak}\ \emph {et~al.}(2013)\citenamefont {Roszak},
	\citenamefont {Mazurek},\ and\ \citenamefont {Horodecki}}]{roszak13}%
\BibitemOpen
\bibfield  {author} {\bibinfo {author} {\bibfnamefont {K.}~\bibnamefont
		{Roszak}}, \bibinfo {author} {\bibfnamefont {P.}~\bibnamefont {Mazurek}}, \
	and\ \bibinfo {author} {\bibfnamefont {P.}~\bibnamefont {Horodecki}},\
}\bibfield  {title} {\enquote {\bibinfo {title} {Anomalous decay of quantum
		correlations of quantum dot qubits},}\ }\href {\doibase
10.1103/PhysRevA.87.062308} {\bibfield  {journal} {\bibinfo  {journal} {Phys.
		Rev. A}\ }\textbf {\bibinfo {volume} {87}},\ \bibinfo {pages} {062308}
(\bibinfo {year} {2013})}\BibitemShut {NoStop}%
\bibitem [{\citenamefont {Mazurek}\ \emph
	{et~al.}(2014{\natexlab{a}})\citenamefont {Mazurek}, \citenamefont {Roszak},\
	and\ \citenamefont {Horodecki}}]{mazurek14}%
\BibitemOpen
\bibfield  {author} {\bibinfo {author} {\bibfnamefont {P.}~\bibnamefont
		{Mazurek}}, \bibinfo {author} {\bibfnamefont {K.}~\bibnamefont {Roszak}}, \
	and\ \bibinfo {author} {\bibfnamefont {P.}~\bibnamefont {Horodecki}},\
}\bibfield  {title} {\enquote {\bibinfo {title} {The decay of quantum
		correlations between quantum dot spin qubits and the characteristics of its
		magnetic-field dependence},}\ }\href
{http://stacks.iop.org/0295-5075/107/i=6/a=67004} {\bibfield  {journal}
	{\bibinfo  {journal} {EPL (Europhysics Letters)}\ }\textbf {\bibinfo {volume}
		{107}},\ \bibinfo {pages} {67004} (\bibinfo {year}
	{2014}{\natexlab{a}})}\BibitemShut {NoStop}%
\bibitem [{\citenamefont {Eisert}\ and\ \citenamefont
	{Plenio}(2002)}]{Eisert_PRL02}%
\BibitemOpen
\bibfield  {author} {\bibinfo {author} {\bibfnamefont {Jens}\ \bibnamefont
		{Eisert}}\ and\ \bibinfo {author} {\bibfnamefont {Martin~B.}\ \bibnamefont
		{Plenio}},\ }\bibfield  {title} {\enquote {\bibinfo {title} {Quantum and
			classical correlations in quantum brownian motion},}\ }\href {\doibase
	10.1103/PhysRevLett.89.137902} {\bibfield  {journal} {\bibinfo  {journal}
		{Phys. Rev. Lett.}\ }\textbf {\bibinfo {volume} {89}},\ \bibinfo {pages}
	{137902} (\bibinfo {year} {2002})}\BibitemShut {NoStop}%
\bibitem [{\citenamefont {Hilt}\ and\ \citenamefont {Lutz}(2009)}]{Hilt_PRA09}%
\BibitemOpen
\bibfield  {author} {\bibinfo {author} {\bibfnamefont {Stefanie}\
		\bibnamefont {Hilt}}\ and\ \bibinfo {author} {\bibfnamefont {Eric}\
		\bibnamefont {Lutz}},\ }\bibfield  {title} {\enquote {\bibinfo {title}
		{System-bath entanglement in quantum thermodynamics},}\ }\href {\doibase
	10.1103/PhysRevA.79.010101} {\bibfield  {journal} {\bibinfo  {journal} {Phys.
			Rev. A}\ }\textbf {\bibinfo {volume} {79}},\ \bibinfo {pages} {010101}
	(\bibinfo {year} {2009})}\BibitemShut {NoStop}%
\bibitem [{\citenamefont {Maziero}\ \emph {et~al.}(2010)\citenamefont
	{Maziero}, \citenamefont {Werlang}, \citenamefont {Fanchini}, \citenamefont
	{C\'eleri},\ and\ \citenamefont {Serra}}]{Maziero_PRA10}%
\BibitemOpen
\bibfield  {author} {\bibinfo {author} {\bibfnamefont {J.}~\bibnamefont
		{Maziero}}, \bibinfo {author} {\bibfnamefont {T.}~\bibnamefont {Werlang}},
	\bibinfo {author} {\bibfnamefont {F.~F.}\ \bibnamefont {Fanchini}}, \bibinfo
	{author} {\bibfnamefont {L.~C.}\ \bibnamefont {C\'eleri}}, \ and\ \bibinfo
	{author} {\bibfnamefont {R.~M.}\ \bibnamefont {Serra}},\ }\bibfield  {title}
{\enquote {\bibinfo {title} {System-reservoir dynamics of quantum and
			classical correlations},}\ }\href {\doibase 10.1103/PhysRevA.81.022116}
{\bibfield  {journal} {\bibinfo  {journal} {Phys. Rev. A}\ }\textbf {\bibinfo
		{volume} {81}},\ \bibinfo {pages} {022116} (\bibinfo {year}
	{2010})}\BibitemShut {NoStop}%
\bibitem [{\citenamefont {Pernice}\ and\ \citenamefont
	{Strunz}(2011)}]{Pernice_PRA11}%
\BibitemOpen
\bibfield  {author} {\bibinfo {author} {\bibfnamefont {A.}~\bibnamefont
		{Pernice}}\ and\ \bibinfo {author} {\bibfnamefont {Walter~T.}\ \bibnamefont
		{Strunz}},\ }\bibfield  {title} {\enquote {\bibinfo {title} {Decoherence and
			the nature of system-environment correlations},}\ }\href {\doibase
	10.1103/PhysRevA.84.062121} {\bibfield  {journal} {\bibinfo  {journal} {Phys.
			Rev. A}\ }\textbf {\bibinfo {volume} {84}},\ \bibinfo {pages} {062121}
	(\bibinfo {year} {2011})}\BibitemShut {NoStop}%
\bibitem [{\citenamefont {Roszak}\ and\ \citenamefont
	{Cywi\ifmmode~\acute{n}\else \'{n}\fi{}ski}(2015)}]{roszak15a}%
\BibitemOpen
\bibfield  {author} {\bibinfo {author} {\bibfnamefont {Katarzyna}\
		\bibnamefont {Roszak}}\ and\ \bibinfo {author} {\bibfnamefont {\L{}ukasz}\
		\bibnamefont {Cywi\ifmmode~\acute{n}\else \'{n}\fi{}ski}},\ }\bibfield
{title} {\enquote {\bibinfo {title} {Characterization and measurement of
			qubit-environment-entanglement generation during pure dephasing},}\ }\href
{\doibase 10.1103/PhysRevA.92.032310} {\bibfield  {journal} {\bibinfo
		{journal} {Phys. Rev. A}\ }\textbf {\bibinfo {volume} {92}},\ \bibinfo
	{pages} {032310} (\bibinfo {year} {2015})}\BibitemShut {NoStop}%
\bibitem [{\citenamefont {Pozzobom}\ and\ \citenamefont
	{Maziero}(2017)}]{Pozzobom_AP17}%
\BibitemOpen
\bibfield  {author} {\bibinfo {author} {\bibfnamefont {Mauro~B.}\
		\bibnamefont {Pozzobom}}\ and\ \bibinfo {author} {\bibfnamefont {Jonas}\
		\bibnamefont {Maziero}},\ }\bibfield  {title} {\enquote {\bibinfo {title}
		{Environment-induced quantum coherence spreading of a qubit},}\ }\href
{\doibase https://doi.org/10.1016/j.aop.2016.12.031} {\bibfield  {journal}
	{\bibinfo  {journal} {Annals of Physics}\ }\textbf {\bibinfo {volume}
		{377}},\ \bibinfo {pages} {243 -- 255} (\bibinfo {year} {2017})}\BibitemShut
{NoStop}%
\bibitem [{\citenamefont {{Rodr{\'i}guez-Rosario}}\ \emph
	{et~al.}(2008)\citenamefont {{Rodr{\'i}guez-Rosario}}, \citenamefont {Modi},
	\citenamefont {Kuah}, \citenamefont {Shaji},\ and\ \citenamefont
	{Sudarshan}}]{Rodriguez_JPA08}%
\BibitemOpen
\bibfield  {author} {\bibinfo {author} {\bibfnamefont {C{\'es}ar~A.}\
		\bibnamefont {{Rodr{\'i}guez-Rosario}}}, \bibinfo {author} {\bibfnamefont
		{Kavan}\ \bibnamefont {Modi}}, \bibinfo {author} {\bibfnamefont {{Aik-meng}}\
		\bibnamefont {Kuah}}, \bibinfo {author} {\bibfnamefont {Anil}\ \bibnamefont
		{Shaji}}, \ and\ \bibinfo {author} {\bibfnamefont {E.~C.~G.}\ \bibnamefont
		{Sudarshan}},\ }\bibfield  {title} {\enquote {\bibinfo {title} {Completely
			positive maps and classical correlations},}\ }\href {\doibase
	10.1088/1751-8113/41/20/205301} {\bibfield  {journal} {\bibinfo  {journal}
		{J.~Phys.~A: Math.~Theor.}\ }\textbf {\bibinfo {volume} {41}},\ \bibinfo
	{pages} {205301} (\bibinfo {year} {2008})}\BibitemShut {NoStop}%
\bibitem [{\citenamefont {Huang}\ \emph {et~al.}(2011)\citenamefont {Huang},
	\citenamefont {Wang},\ and\ \citenamefont {Zhu}}]{huang11}%
\BibitemOpen
\bibfield  {author} {\bibinfo {author} {\bibfnamefont {Jie-Hui}\ \bibnamefont
		{Huang}}, \bibinfo {author} {\bibfnamefont {Lei}\ \bibnamefont {Wang}}, \
	and\ \bibinfo {author} {\bibfnamefont {Shi-Yao}\ \bibnamefont {Zhu}},\
}\bibfield  {title} {\enquote {\bibinfo {title} {A new criterion for zero
		quantum discord},}\ }\href {http://stacks.iop.org/1367-2630/13/i=6/a=063045}
{\bibfield  {journal} {\bibinfo  {journal} {New Journal of Physics}\ }\textbf
	{\bibinfo {volume} {13}},\ \bibinfo {pages} {063045} (\bibinfo {year}
	{2011})}\BibitemShut {NoStop}%
\bibitem [{\citenamefont {Huang}(2014)}]{Huang_NJP14}%
\BibitemOpen
\bibfield  {author} {\bibinfo {author} {\bibfnamefont {Yichen}\ \bibnamefont
		{Huang}},\ }\bibfield  {title} {\enquote {\bibinfo {title} {Computing quantum
			discord is np-complete},}\ }\href
{http://stacks.iop.org/1367-2630/16/i=3/a=033027} {\bibfield  {journal}
	{\bibinfo  {journal} {New Journal of Physics}\ }\textbf {\bibinfo {volume}
		{16}},\ \bibinfo {pages} {033027} (\bibinfo {year} {2014})}\BibitemShut
{NoStop}%
\bibitem [{\citenamefont {Wootters}(1998)}]{wootters98}%
\BibitemOpen
\bibfield  {author} {\bibinfo {author} {\bibfnamefont {William~K.}\
		\bibnamefont {Wootters}},\ }\bibfield  {title} {\enquote {\bibinfo {title}
		{Entanglement of formation of an arbitrary state of two qubits},}\
}\href@noop {} {\ \textbf {\bibinfo {volume} {80}},\ \bibinfo {pages} {2245}
(\bibinfo {year} {1998})}\BibitemShut {NoStop}%
\bibitem [{\citenamefont {Vedral}\ and\ \citenamefont
	{Plenio}(1998)}]{vedral98}%
\BibitemOpen
\bibfield  {author} {\bibinfo {author} {\bibfnamefont {V.}~\bibnamefont
		{Vedral}}\ and\ \bibinfo {author} {\bibfnamefont {M.~B.}\ \bibnamefont
		{Plenio}},\ }\bibfield  {title} {\enquote {\bibinfo {title} {Entanglement
			measures and purification procdures},}\ }\href@noop {} {\bibfield  {journal}
	{\bibinfo  {journal} {Phys. Rev. A}\ }\textbf {\bibinfo {volume} {57}},\
	\bibinfo {pages} {1619} (\bibinfo {year} {1998})}\BibitemShut {NoStop}%
\bibitem [{\citenamefont {Horodecki}(2001)}]{horodecki01}%
\BibitemOpen
\bibfield  {author} {\bibinfo {author} {\bibfnamefont {Micha{\l}}\
		\bibnamefont {Horodecki}},\ }\bibfield  {title} {\enquote {\bibinfo {title}
		{Entanglement measures},}\ }\href@noop {} {\bibfield  {journal} {\bibinfo
		{journal} {Quant. Inf. Comp.}\ }\textbf {\bibinfo {volume} {1}},\ \bibinfo
	{pages} {3} (\bibinfo {year} {2001})}\BibitemShut {NoStop}%
\bibitem [{\citenamefont {Luo}\ and\ \citenamefont {Fu}(2010)}]{luo10}%
\BibitemOpen
\bibfield  {author} {\bibinfo {author} {\bibfnamefont {Shunlong}\
		\bibnamefont {Luo}}\ and\ \bibinfo {author} {\bibfnamefont {Shuangshuang}\
		\bibnamefont {Fu}},\ }\bibfield  {title} {\enquote {\bibinfo {title}
		{Geometric measure of quantum discord},}\ }\href {\doibase
	10.1103/PhysRevA.82.034302} {\bibfield  {journal} {\bibinfo  {journal} {Phys.
			Rev. A}\ }\textbf {\bibinfo {volume} {82}},\ \bibinfo {pages} {034302}
	(\bibinfo {year} {2010})}\BibitemShut {NoStop}%
\bibitem [{\citenamefont {Nakano}\ \emph {et~al.}(2013)\citenamefont {Nakano},
	\citenamefont {Piani},\ and\ \citenamefont {Adesso}}]{nakano13}%
\BibitemOpen
\bibfield  {author} {\bibinfo {author} {\bibfnamefont {Takafumi}\
		\bibnamefont {Nakano}}, \bibinfo {author} {\bibfnamefont {Marco}\
		\bibnamefont {Piani}}, \ and\ \bibinfo {author} {\bibfnamefont {Gerardo}\
		\bibnamefont {Adesso}},\ }\bibfield  {title} {\enquote {\bibinfo {title}
		{Negativity of quantumness and its interpretations},}\ }\href {\doibase
	10.1103/PhysRevA.88.012117} {\bibfield  {journal} {\bibinfo  {journal} {Phys.
			Rev. A}\ }\textbf {\bibinfo {volume} {88}},\ \bibinfo {pages} {012117}
	(\bibinfo {year} {2013})}\BibitemShut {NoStop}%
\bibitem [{\citenamefont {Paula}\ \emph {et~al.}(2013)\citenamefont {Paula},
	\citenamefont {de~Oliveira},\ and\ \citenamefont {Sarandy}}]{paula13}%
\BibitemOpen
\bibfield  {author} {\bibinfo {author} {\bibfnamefont {F.~M.}\ \bibnamefont
		{Paula}}, \bibinfo {author} {\bibfnamefont {Thiago~R.}\ \bibnamefont
		{de~Oliveira}}, \ and\ \bibinfo {author} {\bibfnamefont {M.~S.}\ \bibnamefont
		{Sarandy}},\ }\bibfield  {title} {\enquote {\bibinfo {title} {Geometric
			quantum discord through the schatten 1-norm},}\ }\href {\doibase
	10.1103/PhysRevA.87.064101} {\bibfield  {journal} {\bibinfo  {journal} {Phys.
			Rev. A}\ }\textbf {\bibinfo {volume} {87}},\ \bibinfo {pages} {064101}
	(\bibinfo {year} {2013})}\BibitemShut {NoStop}%
\bibitem [{\citenamefont {Spehner}\ and\ \citenamefont
	{Orszag}(2013)}]{spehner13}%
\BibitemOpen
\bibfield  {author} {\bibinfo {author} {\bibfnamefont {D.}~\bibnamefont
		{Spehner}}\ and\ \bibinfo {author} {\bibfnamefont {M.}~\bibnamefont
		{Orszag}},\ }\bibfield  {title} {\enquote {\bibinfo {title} {Geometric
			quantum discord with bures distance},}\ }\href@noop {} {\bibfield  {journal}
	{\bibinfo  {journal} {New Journal of Physics}\ }\textbf {\bibinfo {volume}
		{15}},\ \bibinfo {pages} {103001} (\bibinfo {year} {2013})}\BibitemShut
{NoStop}%
\bibitem [{\citenamefont {Spehner}\ and\ \citenamefont
	{Orszag}(2014)}]{spehner14}%
\BibitemOpen
\bibfield  {author} {\bibinfo {author} {\bibfnamefont {D.}~\bibnamefont
		{Spehner}}\ and\ \bibinfo {author} {\bibfnamefont {M.}~\bibnamefont
		{Orszag}},\ }\bibfield  {title} {\enquote {\bibinfo {title} {Geometric
			quantum discord with bures distance: the qubit case},}\ }\href@noop {}
{\bibfield  {journal} {\bibinfo  {journal} {J. Phys. A: Math. Theor.}\
	}\textbf {\bibinfo {volume} {47}},\ \bibinfo {pages} {035302} (\bibinfo
	{year} {2014})}\BibitemShut {NoStop}%
\bibitem [{\citenamefont {Miranowicz}\ \emph {et~al.}(2012)\citenamefont
	{Miranowicz}, \citenamefont {Horodecki}, \citenamefont {Chhajlany},
	\citenamefont {Tuziemski},\ and\ \citenamefont {Sperling}}]{miranowicz12}%
\BibitemOpen
\bibfield  {author} {\bibinfo {author} {\bibfnamefont {A.}~\bibnamefont
		{Miranowicz}}, \bibinfo {author} {\bibfnamefont {P.}~\bibnamefont
		{Horodecki}}, \bibinfo {author} {\bibfnamefont {R.~W.}\ \bibnamefont
		{Chhajlany}}, \bibinfo {author} {\bibfnamefont {J.}~\bibnamefont
		{Tuziemski}}, \ and\ \bibinfo {author} {\bibfnamefont {J.}~\bibnamefont
		{Sperling}},\ }\bibfield  {title} {\enquote {\bibinfo {title} {Analytical
			progress on symmetric geometric discord: Measurement-based upper bounds},}\
}\href@noop {} {\bibfield  {journal} {\bibinfo  {journal} {Phys. Rev. A}\
}\textbf {\bibinfo {volume} {86}},\ \bibinfo {pages} {042123} (\bibinfo
{year} {2012})}\BibitemShut {NoStop}%
\bibitem [{\citenamefont {Tufarelli}\ \emph {et~al.}(2013)\citenamefont
	{Tufarelli}, \citenamefont {MacLean}, \citenamefont {Girolami}, \citenamefont
	{Vasile},\ and\ \citenamefont {Adesso}}]{tufarelli13}%
\BibitemOpen
\bibfield  {author} {\bibinfo {author} {\bibfnamefont {Tommaso}\ \bibnamefont
		{Tufarelli}}, \bibinfo {author} {\bibfnamefont {Tom}\ \bibnamefont
		{MacLean}}, \bibinfo {author} {\bibfnamefont {Davide}\ \bibnamefont
		{Girolami}}, \bibinfo {author} {\bibfnamefont {Ruggero}\ \bibnamefont
		{Vasile}}, \ and\ \bibinfo {author} {\bibfnamefont {Gerardo}\ \bibnamefont
		{Adesso}},\ }\bibfield  {title} {\enquote {\bibinfo {title} {The geometric
			approach to quantum correlations: Computability versus reliability},}\
}\href@noop {} {\bibfield  {journal} {\bibinfo  {journal} {J. Phys. A}\
}\textbf {\bibinfo {volume} {46}},\ \bibinfo {pages} {275308} (\bibinfo
{year} {2013})}\BibitemShut {NoStop}%
\bibitem [{\citenamefont {Sza\'nkowski}\ \emph {et~al.}(2017)\citenamefont
	{Sza\'nkowski}, \citenamefont {Ramon}, \citenamefont {Krzywda}, \citenamefont
	{Kwiatkowski},\ and\ \citenamefont {Cywi\'nski}}]{Szankowski_JPCM17}%
\BibitemOpen
\bibfield  {author} {\bibinfo {author} {\bibfnamefont {P.}~\bibnamefont
		{Sza\'nkowski}}, \bibinfo {author} {\bibfnamefont {G.}~\bibnamefont {Ramon}},
	\bibinfo {author} {\bibfnamefont {J.}~\bibnamefont {Krzywda}}, \bibinfo
	{author} {\bibfnamefont {D.}~\bibnamefont {Kwiatkowski}}, \ and\ \bibinfo
	{author} {\bibfnamefont {\L.}\ \bibnamefont {Cywi\'nski}},\ }\bibfield
{title} {\enquote {\bibinfo {title} {Environmental noise spectroscopy with
			qubits subjected to dynamical decoupling},}\ }\href {\doibase
	10.1088/1361-648X/aa7648} {\bibfield  {journal} {\bibinfo  {journal} {J.
			Phys.:Condens. Matter}\ }\textbf {\bibinfo {volume} {29}},\ \bibinfo {pages}
	{333001} (\bibinfo {year} {2017})}\BibitemShut {NoStop}%
\bibitem [{\citenamefont {Monz}\ \emph {et~al.}(2011)\citenamefont {Monz},
	\citenamefont {Schindler}, \citenamefont {Barreiro}, \citenamefont {Chwalla},
	\citenamefont {Nigg}, \citenamefont {Coish}, \citenamefont {Harlander},
	\citenamefont {H\"ansel}, \citenamefont {Hennrich},\ and\ \citenamefont
	{Blatt}}]{Monz_PRL11}%
\BibitemOpen
\bibfield  {author} {\bibinfo {author} {\bibfnamefont {Thomas}\ \bibnamefont
		{Monz}}, \bibinfo {author} {\bibfnamefont {Philipp}\ \bibnamefont
		{Schindler}}, \bibinfo {author} {\bibfnamefont {Julio~T.}\ \bibnamefont
		{Barreiro}}, \bibinfo {author} {\bibfnamefont {Michael}\ \bibnamefont
		{Chwalla}}, \bibinfo {author} {\bibfnamefont {Daniel}\ \bibnamefont {Nigg}},
	\bibinfo {author} {\bibfnamefont {William~A.}\ \bibnamefont {Coish}},
	\bibinfo {author} {\bibfnamefont {Maximilian}\ \bibnamefont {Harlander}},
	\bibinfo {author} {\bibfnamefont {Wolfgang}\ \bibnamefont {H\"ansel}},
	\bibinfo {author} {\bibfnamefont {Markus}\ \bibnamefont {Hennrich}}, \ and\
	\bibinfo {author} {\bibfnamefont {Rainer}\ \bibnamefont {Blatt}},\ }\bibfield
{title} {\enquote {\bibinfo {title} {14-qubit entanglement: Creation and
			coherence},}\ }\href {\doibase 10.1103/PhysRevLett.106.130506} {\bibfield
	{journal} {\bibinfo  {journal} {Phys. Rev. Lett.}\ }\textbf {\bibinfo
		{volume} {106}},\ \bibinfo {pages} {130506} (\bibinfo {year}
	{2011})}\BibitemShut {NoStop}%
\bibitem [{\citenamefont {Roszak}\ and\ \citenamefont
	{Machnikowski}(2006)}]{roszak06b}%
\BibitemOpen
\bibfield  {author} {\bibinfo {author} {\bibfnamefont {K.}~\bibnamefont
		{Roszak}}\ and\ \bibinfo {author} {\bibfnamefont {P.}~\bibnamefont
		{Machnikowski}},\ }\bibfield  {title} {\enquote {\bibinfo {title} {{``Which
				path''} decoherence in quantum dot experiments},}\ }\href@noop {} {\ \textbf
	{\bibinfo {volume} {351}},\ \bibinfo {pages} {251--256} (\bibinfo {year}
	{2006})}\BibitemShut {NoStop}%
\bibitem [{\citenamefont {Krzywda}\ and\ \citenamefont
	{Roszak}(2016)}]{Krzywda_SR16}%
\BibitemOpen
\bibfield  {author} {\bibinfo {author} {\bibfnamefont {Jan}\ \bibnamefont
		{Krzywda}}\ and\ \bibinfo {author} {\bibfnamefont {Katarzyna}\ \bibnamefont
		{Roszak}},\ }\bibfield  {title} {\enquote {\bibinfo {title} {Complete
			disentanglement by partial pure dephasing},}\ }\href {\doibase
	10.1038/srep23753} {\bibfield  {journal} {\bibinfo  {journal} {Sci. Rep.}\
	}\textbf {\bibinfo {volume} {6}},\ \bibinfo {pages} {23753} (\bibinfo {year}
	{2016})}\BibitemShut {NoStop}%
\bibitem [{\citenamefont {Salamon}\ and\ \citenamefont
	{Roszak}(2017)}]{Salamon_PRA17}%
\BibitemOpen
\bibfield  {author} {\bibinfo {author} {\bibfnamefont {Tymoteusz}\
		\bibnamefont {Salamon}}\ and\ \bibinfo {author} {\bibfnamefont {Katarzyna}\
		\bibnamefont {Roszak}},\ }\bibfield  {title} {\enquote {\bibinfo {title}
		{Entanglement generation between a charge qubit and its bosonic environment
			during pure dephasing: Dependence on the environment size},}\ }\href
{\doibase 10.1103/PhysRevA.96.032333} {\bibfield  {journal} {\bibinfo
		{journal} {Phys. Rev. A}\ }\textbf {\bibinfo {volume} {96}},\ \bibinfo
	{pages} {032333} (\bibinfo {year} {2017})}\BibitemShut {NoStop}%
\bibitem [{\citenamefont {Paz-Silva}\ \emph {et~al.}(2017)\citenamefont
	{Paz-Silva}, \citenamefont {Norris},\ and\ \citenamefont
	{Viola}}]{Paz_PRA17}%
\BibitemOpen
\bibfield  {author} {\bibinfo {author} {\bibfnamefont {Gerardo~A.}\
		\bibnamefont {Paz-Silva}}, \bibinfo {author} {\bibfnamefont {Leigh~M.}\
		\bibnamefont {Norris}}, \ and\ \bibinfo {author} {\bibfnamefont {Lorenza}\
		\bibnamefont {Viola}},\ }\bibfield  {title} {\enquote {\bibinfo {title}
		{Multiqubit spectroscopy of gaussian quantum noise},}\ }\href {\doibase
	10.1103/PhysRevA.95.022121} {\bibfield  {journal} {\bibinfo  {journal} {Phys.
			Rev. A}\ }\textbf {\bibinfo {volume} {95}},\ \bibinfo {pages} {022121}
	(\bibinfo {year} {2017})}\BibitemShut {NoStop}%
\bibitem [{\citenamefont {Cywi{\'n}ski}\ \emph {et~al.}(2009)\citenamefont
	{Cywi{\'n}ski}, \citenamefont {Witzel},\ and\ \citenamefont {{Das
			Sarma}}}]{Cywinski_PRB09}%
\BibitemOpen
\bibfield  {author} {\bibinfo {author} {\bibfnamefont {{\L}ukasz}\
		\bibnamefont {Cywi{\'n}ski}}, \bibinfo {author} {\bibfnamefont {Wayne~M.}\
		\bibnamefont {Witzel}}, \ and\ \bibinfo {author} {\bibfnamefont
		{S.}~\bibnamefont {{Das Sarma}}},\ }\bibfield  {title} {\enquote {\bibinfo
		{title} {Pure quantum dephasing of a solid-state electron spin qubit in a
			large nuclear spin bath coupled by long-range hyperfine-mediated
			interaction},}\ }\href {\doibase 10.1103/PhysRevB.79.245314} {\bibfield
	{journal} {\bibinfo  {journal} {Phys.\ Rev.\ B}\ }\textbf {\bibinfo {volume}
		{79}},\ \bibinfo {pages} {245314} (\bibinfo {year} {2009})}\BibitemShut
{NoStop}%
\bibitem [{\citenamefont {Yang}\ \emph {et~al.}(2017)\citenamefont {Yang},
	\citenamefont {Ma}, ,\ and\ \citenamefont {Liu}}]{Yang_RPP17}%
\BibitemOpen
\bibfield  {author} {\bibinfo {author} {\bibfnamefont {Wen}\ \bibnamefont
		{Yang}}, \bibinfo {author} {\bibfnamefont {Wen-Long}\ \bibnamefont {Ma}}, , \
	and\ \bibinfo {author} {\bibfnamefont {Ren-Bao}\ \bibnamefont {Liu}},\
}\bibfield  {title} {\enquote {\bibinfo {title} {Quantum many-body theory for
		electron spin decoherence in nanoscale nuclear spin baths},}\ }\href
{\doibase 10.1088/0034-4885/80/1/016001} {\bibfield  {journal} {\bibinfo
		{journal} {Rep. Prog. Phys.}\ }\textbf {\bibinfo {volume} {80}},\ \bibinfo
	{pages} {016001} (\bibinfo {year} {2017})}\BibitemShut {NoStop}%
\bibitem [{\citenamefont {Zhao}\ \emph {et~al.}(2012)\citenamefont {Zhao},
	\citenamefont {Ho},\ and\ \citenamefont {Liu}}]{Zhao_PRB12}%
\BibitemOpen
\bibfield  {author} {\bibinfo {author} {\bibfnamefont {Nan}\ \bibnamefont
		{Zhao}}, \bibinfo {author} {\bibfnamefont {Sai-Wah}\ \bibnamefont {Ho}}, \
	and\ \bibinfo {author} {\bibfnamefont {Ren-Bao}\ \bibnamefont {Liu}},\
}\bibfield  {title} {\enquote {\bibinfo {title} {Decoherence and dynamical
		decoupling control of nitrogen vacancy center electron spins in nuclear spin
		baths},}\ }\href {\doibase 10.1103/PhysRevB.85.115303} {\bibfield  {journal}
{\bibinfo  {journal} {Phys. Rev. B}\ }\textbf {\bibinfo {volume} {85}},\
\bibinfo {pages} {115303} (\bibinfo {year} {2012})}\BibitemShut {NoStop}%
\bibitem [{\citenamefont {{de Lange}}\ \emph {et~al.}(2010)\citenamefont {{de
			Lange}}, \citenamefont {Wang}, \citenamefont {Rist{\`e}}, \citenamefont
	{Dobrovitski},\ and\ \citenamefont {Hanson}}]{deLange_Science10}%
\BibitemOpen
\bibfield  {author} {\bibinfo {author} {\bibfnamefont {G.}~\bibnamefont {{de
				Lange}}}, \bibinfo {author} {\bibfnamefont {Z.~H.}\ \bibnamefont {Wang}},
	\bibinfo {author} {\bibfnamefont {D.}~\bibnamefont {Rist{\`e}}}, \bibinfo
	{author} {\bibfnamefont {V.~V.}\ \bibnamefont {Dobrovitski}}, \ and\ \bibinfo
	{author} {\bibfnamefont {R.}~\bibnamefont {Hanson}},\ }\bibfield  {title}
{\enquote {\bibinfo {title} {Universal dynamical decoupling of a single
			solid-state spin from a spin bath},}\ }\href {\doibase
	10.1126/science.1192739} {\bibfield  {journal} {\bibinfo  {journal}
		{Science}\ }\textbf {\bibinfo {volume} {330}},\ \bibinfo {pages} {60}
	(\bibinfo {year} {2010})}\BibitemShut {NoStop}%
\bibitem [{\citenamefont {Witzel}\ \emph {et~al.}(2012)\citenamefont {Witzel},
	\citenamefont {Carroll}, \citenamefont {Cywi{\'n}ski},\ and\ \citenamefont
	{Das~Sarma}}]{Witzel_PRB12}%
\BibitemOpen
\bibfield  {author} {\bibinfo {author} {\bibfnamefont {Wayne~M.}\
		\bibnamefont {Witzel}}, \bibinfo {author} {\bibfnamefont {Malcolm~S.}\
		\bibnamefont {Carroll}}, \bibinfo {author} {\bibfnamefont {{\L}ukasz}\
		\bibnamefont {Cywi{\'n}ski}}, \ and\ \bibinfo {author} {\bibfnamefont
		{S.}~\bibnamefont {Das~Sarma}},\ }\bibfield  {title} {\enquote {\bibinfo
		{title} {Quantum decoherence of the central spin in a sparse system of
			dipolar coupled spins},}\ }\href {\doibase 10.1103/PhysRevB.86.035452}
{\bibfield  {journal} {\bibinfo  {journal} {Phys.\ Rev.\ B}\ }\textbf
	{\bibinfo {volume} {86}},\ \bibinfo {pages} {035452} (\bibinfo {year}
	{2012})}\BibitemShut {NoStop}%
\bibitem [{\citenamefont {Zhao}\ \emph {et~al.}(2011)\citenamefont {Zhao},
	\citenamefont {Wang},\ and\ \citenamefont {Liu}}]{Zhao_PRL11}%
\BibitemOpen
\bibfield  {author} {\bibinfo {author} {\bibfnamefont {Nan}\ \bibnamefont
		{Zhao}}, \bibinfo {author} {\bibfnamefont {Zhen-Yu}\ \bibnamefont {Wang}}, \
	and\ \bibinfo {author} {\bibfnamefont {Ren-Bao}\ \bibnamefont {Liu}},\
}\bibfield  {title} {\enquote {\bibinfo {title} {Anomalous decoherence effect
		in a quantum bath},}\ }\href {\doibase 10.1103/PhysRevLett.106.217205}
{\bibfield  {journal} {\bibinfo  {journal} {Phys.\ Rev.\ Lett.}\ }\textbf
	{\bibinfo {volume} {106}},\ \bibinfo {pages} {217205} (\bibinfo {year}
	{2011})}\BibitemShut {NoStop}%
\bibitem [{\citenamefont {Hassan}\ and\ \citenamefont {Joag}(2013)}]{hassan13}%
\BibitemOpen
\bibfield  {author} {\bibinfo {author} {\bibfnamefont {Ali Saif~M.}\
		\bibnamefont {Hassan}}\ and\ \bibinfo {author} {\bibfnamefont {Pramod~S.}\
		\bibnamefont {Joag}},\ }\bibfield  {title} {\enquote {\bibinfo {title}
		{Invariance of quantum correlations under local channel for a bipartite
			quantum state},}\ }\href {http://stacks.iop.org/0295-5075/103/i=1/a=10004}
{\bibfield  {journal} {\bibinfo  {journal} {EPL (Europhysics Letters)}\
	}\textbf {\bibinfo {volume} {103}},\ \bibinfo {pages} {10004} (\bibinfo
	{year} {2013})}\BibitemShut {NoStop}%
\bibitem [{\citenamefont {{\.Z}urek}(2003)}]{Zurek_RMP03}%
\BibitemOpen
\bibfield  {author} {\bibinfo {author} {\bibfnamefont {Wojciech~Hubert}\
		\bibnamefont {{\.Z}urek}},\ }\bibfield  {title} {\enquote {\bibinfo {title}
		{Decoherence, einselection, and the quantum origins of the classical},}\
}\href {\doibase 10.1103/RevModPhys.75.715} {\bibfield  {journal} {\bibinfo
	{journal} {Rev.\ Mod.\ Phys.}\ }\textbf {\bibinfo {volume} {75}},\ \bibinfo
{pages} {715} (\bibinfo {year} {2003})}\BibitemShut {NoStop}%
\bibitem [{\citenamefont {Buscemi}(2014)}]{Buscemi_PRL14}%
\BibitemOpen
\bibfield  {author} {\bibinfo {author} {\bibfnamefont {Francesco}\
		\bibnamefont {Buscemi}},\ }\bibfield  {title} {\enquote {\bibinfo {title}
		{Complete positivity, markovianity, and the quantum data-processing
			inequality, in the presence of initial system-environment correlations},}\
}\href {\doibase 10.1103/PhysRevLett.113.140502} {\bibfield  {journal}
{\bibinfo  {journal} {Phys. Rev. Lett.}\ }\textbf {\bibinfo {volume} {113}},\
\bibinfo {pages} {140502} (\bibinfo {year} {2014})}\BibitemShut {NoStop}%
\bibitem [{\citenamefont {Vacchini}\ and\ \citenamefont
	{Amato}(2016)}]{Vacchini_SR16}%
\BibitemOpen
\bibfield  {author} {\bibinfo {author} {\bibfnamefont {Bassano}\ \bibnamefont
		{Vacchini}}\ and\ \bibinfo {author} {\bibfnamefont {Giulio}\ \bibnamefont
		{Amato}},\ }\bibfield  {title} {\enquote {\bibinfo {title} {Reduced dynamical
			maps in the presence of initial correlations},}\ }\href {\doibase
	10.1038/srep37328} {\bibfield  {journal} {\bibinfo  {journal} {Sci.~Rep.}\
	}\textbf {\bibinfo {volume} {6}},\ \bibinfo {pages} {37328} (\bibinfo {year}
	{2016})}\BibitemShut {NoStop}%
\bibitem [{\citenamefont {Dominy}\ \emph {et~al.}(2016)\citenamefont {Dominy},
	\citenamefont {Shabani},\ and\ \citenamefont
	{Lidar}}]{Dominy_framework_CP_QIP16}%
\BibitemOpen
\bibfield  {author} {\bibinfo {author} {\bibfnamefont {Jason~M.}\
		\bibnamefont {Dominy}}, \bibinfo {author} {\bibfnamefont {Alireza}\
		\bibnamefont {Shabani}}, \ and\ \bibinfo {author} {\bibfnamefont {Daniel~A.}\
		\bibnamefont {Lidar}},\ }\bibfield  {title} {\enquote {\bibinfo {title} {A
			general framework for complete positivity},}\ }\href {\doibase
	10.1007/s11128-015-1148-0} {\bibfield  {journal} {\bibinfo  {journal}
		{Quantum Info. Process.}\ }\textbf {\bibinfo {volume} {15}},\ \bibinfo
	{pages} {465} (\bibinfo {year} {2016})}\BibitemShut {NoStop}%
\bibitem [{\citenamefont {Dominy}\ and\ \citenamefont
	{Lidar}(2016)}]{Dominy_beyond_CP_QIP16}%
\BibitemOpen
\bibfield  {author} {\bibinfo {author} {\bibfnamefont {Jason~M.}\
		\bibnamefont {Dominy}}\ and\ \bibinfo {author} {\bibfnamefont {Daniel~A.}\
		\bibnamefont {Lidar}},\ }\bibfield  {title} {\enquote {\bibinfo {title}
		{Beyond complete positivity},}\ }\href {\doibase 10.1007/s11128-015-1228-1}
{\bibfield  {journal} {\bibinfo  {journal} {Quantum Info. Process.}\ }\textbf
	{\bibinfo {volume} {15}},\ \bibinfo {pages} {1349} (\bibinfo {year}
	{2016})}\BibitemShut {NoStop}%
\bibitem [{\citenamefont {Yu}\ and\ \citenamefont {Eberly}(2007)}]{Yu_QIC07}%
\BibitemOpen
\bibfield  {author} {\bibinfo {author} {\bibfnamefont {Ting}\ \bibnamefont
		{Yu}}\ and\ \bibinfo {author} {\bibfnamefont {J.~H.}\ \bibnamefont
		{Eberly}},\ }\bibfield  {title} {\enquote {\bibinfo {title} {Evolution from
			entanglement to decoherence},}\ }\href@noop {} {\bibfield  {journal}
	{\bibinfo  {journal} {Quant.~Info.~Comp.}\ }\textbf {\bibinfo {volume} {7}},\
	\bibinfo {pages} {459} (\bibinfo {year} {2007})}\BibitemShut {NoStop}%
\bibitem [{\citenamefont {Mazurek}\ \emph
	{et~al.}(2014{\natexlab{b}})\citenamefont {Mazurek}, \citenamefont {Roszak},
	\citenamefont {Chhajlany},\ and\ \citenamefont {Horodecki}}]{mazurek14a}%
\BibitemOpen
\bibfield  {author} {\bibinfo {author} {\bibfnamefont {Pawe\l{}}\
		\bibnamefont {Mazurek}}, \bibinfo {author} {\bibfnamefont {Katarzyna}\
		\bibnamefont {Roszak}}, \bibinfo {author} {\bibfnamefont {Ravindra~W.}\
		\bibnamefont {Chhajlany}}, \ and\ \bibinfo {author} {\bibfnamefont
		{Pawe\l{}}\ \bibnamefont {Horodecki}},\ }\bibfield  {title} {\enquote
	{\bibinfo {title} {Sensitivity of entanglement decay of quantum-dot spin
			qubits to the external magnetic field},}\ }\href {\doibase
	10.1103/PhysRevA.89.062318} {\bibfield  {journal} {\bibinfo  {journal} {Phys.
			Rev. A}\ }\textbf {\bibinfo {volume} {89}},\ \bibinfo {pages} {062318}
	(\bibinfo {year} {2014}{\natexlab{b}})}\BibitemShut {NoStop}%
\bibitem [{\citenamefont {Sza{\'n}kowski}\ \emph {et~al.}(2015)\citenamefont
	{Sza{\'n}kowski}, \citenamefont {Trippenbach}, \citenamefont {Cywi{\'n}ski},\
	and\ \citenamefont {Band}}]{Szankowski_QIP15}%
\BibitemOpen
\bibfield  {author} {\bibinfo {author} {\bibfnamefont {Piotr}\ \bibnamefont
		{Sza{\'n}kowski}}, \bibinfo {author} {\bibfnamefont {Marek}\ \bibnamefont
		{Trippenbach}}, \bibinfo {author} {\bibfnamefont {{\L}ukasz}\ \bibnamefont
		{Cywi{\'n}ski}}, \ and\ \bibinfo {author} {\bibfnamefont {Y.~B.}\
		\bibnamefont {Band}},\ }\bibfield  {title} {\enquote {\bibinfo {title} {The
			dynamics of two entangled qubits exposed to classical noise: role of spatial
			and temporal noise correlations},}\ }\href {\doibase
	10.1007/s11128-015-1044-7} {\bibfield  {journal} {\bibinfo  {journal}
		{Quantum Inf.~Process.}\ }\textbf {\bibinfo {volume} {14}},\ \bibinfo {pages}
	{3367} (\bibinfo {year} {2015})}\BibitemShut {NoStop}%
\bibitem [{\citenamefont {Bragar}\ and\ \citenamefont
	{Cywi{\'n}ski}(2015)}]{Bragar_PRB15}%
\BibitemOpen
\bibfield  {author} {\bibinfo {author} {\bibfnamefont {Igor}\ \bibnamefont
		{Bragar}}\ and\ \bibinfo {author} {\bibfnamefont {{\L}ukasz}\ \bibnamefont
		{Cywi{\'n}ski}},\ }\bibfield  {title} {\enquote {\bibinfo {title} {Dynamics
			of entanglement of two electron spins interacting with nuclear spin baths in
			quantum dots},}\ }\href {\doibase 10.1103/PhysRevB.91.155310} {\bibfield
	{journal} {\bibinfo  {journal} {Phys. Rev. B}\ }\textbf {\bibinfo {volume}
		{91}},\ \bibinfo {pages} {155310} (\bibinfo {year} {2015})}\BibitemShut
{NoStop}%
\bibitem [{\citenamefont {Aolita}\ \emph {et~al.}(2015)\citenamefont {Aolita},
	\citenamefont {{de Melo}},\ and\ \citenamefont {Davidovich}}]{Aolita_RPP15}%
\BibitemOpen
\bibfield  {author} {\bibinfo {author} {\bibfnamefont {Leandro}\ \bibnamefont
		{Aolita}}, \bibinfo {author} {\bibfnamefont {Fernando}\ \bibnamefont {{de
				Melo}}}, \ and\ \bibinfo {author} {\bibfnamefont {Luiz}\ \bibnamefont
		{Davidovich}},\ }\bibfield  {title} {\enquote {\bibinfo {title} {Open-system
			dynamics of entanglement},}\ }\href {\doibase 10.1088/0034-4885/78/4/042001}
{\bibfield  {journal} {\bibinfo  {journal} {Rep. Prog. Phys.}\ }\textbf
	{\bibinfo {volume} {78}},\ \bibinfo {pages} {042001} (\bibinfo {year}
	{2015})}\BibitemShut {NoStop}%
\bibitem [{\citenamefont {Chitambar}\ and\ \citenamefont
	{Duan}(2009)}]{Chitambar_PRL09}%
\BibitemOpen
\bibfield  {author} {\bibinfo {author} {\bibfnamefont {Eric}\ \bibnamefont
		{Chitambar}}\ and\ \bibinfo {author} {\bibfnamefont {Runyao}\ \bibnamefont
		{Duan}},\ }\bibfield  {title} {\enquote {\bibinfo {title} {Nonlocal
			entanglement transformations achievable by separable operations},}\ }\href
{\doibase 10.1103/PhysRevLett.103.110502} {\bibfield  {journal} {\bibinfo
		{journal} {Phys. Rev. Lett.}\ }\textbf {\bibinfo {volume} {103}},\ \bibinfo
	{pages} {110502} (\bibinfo {year} {2009})}\BibitemShut {NoStop}%
\bibitem [{\citenamefont {Gheorghiu}\ and\ \citenamefont
	{Griffiths}(2008)}]{Gheorghiu_PRA08}%
\BibitemOpen
\bibfield  {author} {\bibinfo {author} {\bibfnamefont {Vlad}\ \bibnamefont
		{Gheorghiu}}\ and\ \bibinfo {author} {\bibfnamefont {Robert~B.}\ \bibnamefont
		{Griffiths}},\ }\bibfield  {title} {\enquote {\bibinfo {title} {Separable
			operations on pure states},}\ }\href {\doibase 10.1103/PhysRevA.78.020304}
{\bibfield  {journal} {\bibinfo  {journal} {Phys. Rev. A}\ }\textbf {\bibinfo
		{volume} {78}},\ \bibinfo {pages} {020304} (\bibinfo {year}
	{2008})}\BibitemShut {NoStop}%
\bibitem [{\citenamefont {{D'Arrigo}}\ \emph {et~al.}(2014)\citenamefont
	{{D'Arrigo}}, \citenamefont {{Lo Franco}}, \citenamefont {Benenti},
	\citenamefont {Paladino},\ and\ \citenamefont {Falci}}]{DArrigo_AP14}%
\BibitemOpen
\bibfield  {author} {\bibinfo {author} {\bibfnamefont {A.}~\bibnamefont
		{{D'Arrigo}}}, \bibinfo {author} {\bibfnamefont {R.}~\bibnamefont {{Lo
				Franco}}}, \bibinfo {author} {\bibfnamefont {G.}~\bibnamefont {Benenti}},
	\bibinfo {author} {\bibfnamefont {E.}~\bibnamefont {Paladino}}, \ and\
	\bibinfo {author} {\bibfnamefont {G.}~\bibnamefont {Falci}},\ }\bibfield
{title} {\enquote {\bibinfo {title} {Recovering entanglement by local
			operations},}\ }\href {\doibase 10.1016/j.aop.2014.07.021} {\bibfield
	{journal} {\bibinfo  {journal} {Ann.~Phys.}\ }\textbf {\bibinfo {volume}
		{350}},\ \bibinfo {pages} {211} (\bibinfo {year} {2014})}\BibitemShut
{NoStop}%
\bibitem [{\citenamefont {Krisnanda}\ \emph {et~al.}(2017)\citenamefont
	{Krisnanda}, \citenamefont {Zuppardo}, \citenamefont {Paternostro},\ and\
	\citenamefont {Paterek}}]{krisnanda17}%
\BibitemOpen
\bibfield  {author} {\bibinfo {author} {\bibfnamefont {Tanjung}\ \bibnamefont
		{Krisnanda}}, \bibinfo {author} {\bibfnamefont {Margherita}\ \bibnamefont
		{Zuppardo}}, \bibinfo {author} {\bibfnamefont {Mauro}\ \bibnamefont
		{Paternostro}}, \ and\ \bibinfo {author} {\bibfnamefont {Tomasz}\
		\bibnamefont {Paterek}},\ }\bibfield  {title} {\enquote {\bibinfo {title}
		{Revealing non-classicality of inaccessible objects},}\ }\href@noop {} {\
	(\bibinfo {year} {2017})},\ \bibinfo {note} {arXiv:1607.01140v2
	[quant-ph]}\BibitemShut {NoStop}%
\bibitem [{\citenamefont {Sargolzahi}\ and\ \citenamefont
	{Mirafzali}(2017)}]{sargolzahi17}%
\BibitemOpen
\bibfield  {author} {\bibinfo {author} {\bibfnamefont {Iman}\ \bibnamefont
		{Sargolzahi}}\ and\ \bibinfo {author} {\bibfnamefont {Sayyed~Yahya}\
		\bibnamefont {Mirafzali}},\ }\bibfield  {title} {\enquote {\bibinfo {title}
		{Entanglement increase from local interaction in the absence of initial
			quantum correlation in the environment and between the system and the
			environment},}\ }\href@noop {} {\  (\bibinfo {year} {2017})},\ \bibinfo
{note} {arXiv:1706.05600v2 [quant-ph]}\BibitemShut {NoStop}%
\bibitem [{\citenamefont {Plenio}(2005)}]{plenio05b}%
\BibitemOpen
\bibfield  {author} {\bibinfo {author} {\bibfnamefont {M.~B.}\ \bibnamefont
		{Plenio}},\ }\bibfield  {title} {\enquote {\bibinfo {title} {Logarithmic
			negativity: A full entanglement monotone that is not convex},}\ }\href
{\doibase 10.1103/PhysRevLett.95.090503} {\bibfield  {journal} {\bibinfo
		{journal} {Phys. Rev. Lett.}\ }\textbf {\bibinfo {volume} {95}},\ \bibinfo
	{pages} {090503} (\bibinfo {year} {2005})}\BibitemShut {NoStop}%
\end{thebibliography}
\end{document}